\documentclass[aps,prl,showpacs,english,10pt,a4,nofootinbib,notitlepage,twocolumn,superscriptaddress]{revtex4-1}
\usepackage[figuresright]{rotating}

\usepackage{graphicx}% Include figure files
\usepackage{dcolumn}% Align table columns on decimal point
\usepackage{bm}% bold math
\usepackage{hyperref}% add hypertext capabilities
\usepackage[mathlines]{lineno}% Enable numbering of text and display math
\usepackage{blkarray}
\usepackage{bbold}
\usepackage{epsfig}
\usepackage{hhline}
\usepackage{multirow}

\usepackage{array}
\newcolumntype{L}[1]{>{\raggedright\let\newline\\\arraybackslash\hspace{0pt}}m{#1}}
\newcolumntype{C}[1]{>{\centering\let\newline\\\arraybackslash\hspace{0pt}}m{#1}}
\newcolumntype{R}[1]{>{\raggedleft\let\newline\\\arraybackslash\hspace{0pt}}m{#1}}

\usepackage{bigdelim}
\newcolumntype{K}[1]{>{\centering\arraybackslash}p{#1}}
\newcolumntype{M}[1]{>{\centering\arraybackslash}m{#1}}

\usepackage[utf8]{inputenc}
\usepackage[T1]{fontenc}
\usepackage[english]{babel}  % english language

\usepackage{times}

\usepackage{amssymb,amsmath,amsfonts,amsthm,dsfont}

\usepackage{verbatim}
\usepackage{enumerate}

\graphicspath{{pics/}}
\usepackage{hyperref}
\usepackage{bbm}
\usepackage{booktabs}

\usepackage[caption=false]{subfig}

\usepackage{color}
\definecolor{dred}{rgb}{.8,0.2,.2}
\definecolor{ddred}{rgb}{.8,0.5,.5}
\definecolor{dblue}{rgb}{.2,0.2,.8}
\definecolor{dgreen}{rgb}{.2,0.5,.2}
\newcommand{\bra}[1]{\mbox{$\langle #1|$}}
\newcommand{\ket}[1]{\ensuremath{|#1\rangle}}

\DeclareMathOperator{\Tr}{Tr}

\def\one{{\mathchoice {\rm 1\mskip-4mu l} {\rm 1\mskip-4mu l} {\rm
1\mskip-4.5mu l} {\rm 1\mskip-5mu l}}}
\newcommand{\be}{\begin{equation}}
\newcommand{\ee}{\end{equation}}

\begin{document}
\title{Estimating the coherence of noise in quantum control of a solid-state qubit}
\date{\today}

\author{Guanru Feng}
\affiliation{Institute for Quantum Computing, Waterloo, Ontario, N2L 3G1, Canada}
\affiliation{Department of Physics and Astronomy, University of Waterloo, Waterloo, Ontario, N2L 3G1, Canada}

\author{Joel J. Wallman}
\affiliation{Institute for Quantum Computing, Waterloo, Ontario, N2L 3G1, Canada}
\affiliation{Department of Applied Mathematics, University of Waterloo, Waterloo, Ontario, N2L 3G1, Canada}

\author{Brandon Buonacorsi}
\affiliation{Institute for Quantum Computing, Waterloo, Ontario, N2L 3G1, Canada}
\affiliation{Department of Physics and Astronomy, University of Waterloo, Waterloo, Ontario, N2L 3G1, Canada}

\author{Franklin H. Cho}
\affiliation{Institute for Quantum Computing, Waterloo, Ontario, N2L 3G1, Canada}
\affiliation{Department of Physics and Astronomy, University of Waterloo, Waterloo, Ontario, N2L 3G1, Canada}

\author{Daniel K. Park}
\affiliation{Institute for Quantum Computing, Waterloo, Ontario, N2L 3G1, Canada}
\affiliation{Department of Physics and Astronomy, University of Waterloo, Waterloo, Ontario, N2L 3G1, Canada}
\affiliation{Natural Science Research Institute, Korea Advanced Institute of Science and Technology, Daejon, 34141, South Korea}

\author{Tao Xin} 
\affiliation{Institute for Quantum Computing, Waterloo, Ontario, N2L 3G1, Canada}
\affiliation{Department of Physics and Astronomy, University of Waterloo, Waterloo, Ontario, N2L 3G1, Canada}
\affiliation{Department of Physics, Tsinghua University, Beijing 100084, China}

\author{Dawei Lu}
\affiliation{Institute for Quantum Computing, Waterloo, Ontario, N2L 3G1, Canada}
\affiliation{Department of Physics and Astronomy, University of Waterloo, Waterloo, Ontario, N2L 3G1, Canada}

\author{Jonathan Baugh}
\email{baugh@uwaterloo.ca}
\affiliation{Institute for Quantum Computing, Waterloo, Ontario, N2L 3G1, Canada}
\affiliation{Department of Physics and Astronomy, University of Waterloo, Waterloo, Ontario, N2L 3G1, Canada}
\affiliation{Department of Chemistry, University of Waterloo, Waterloo, Ontario, N2L 3G1, Canada}

\author{Raymond Laflamme}
\email{laflamme@iqc.ca}
\affiliation{Institute for Quantum Computing, Waterloo, Ontario, N2L 3G1, Canada}
\affiliation{Department of Physics and Astronomy, University of Waterloo, Waterloo, Ontario, N2L 3G1, Canada}
\affiliation{Perimeter Institute for Theoretical Physics, Waterloo, Ontario, N2J 2W9, Canada}
\affiliation{Canadian Institute for Advanced Research, Toronto, Ontario M5G 1Z8, Canada}

\begin{abstract}
To exploit a given physical system for quantum information processing, it is 
critical to understand the different types of noise affecting quantum control. 
Distinguishing coherent and incoherent errors is extremely useful as they can 
be reduced in different ways. Coherent errors are generally easier to reduce at 
the hardware level, e.g. by improving calibration, whereas some sources of 
incoherent errors, e.g. $T_2^*$ processes, can be reduced by engineering robust pulses. In this work, we illustrate how purity benchmarking and 
randomized benchmarking can be used together to distinguish between coherent 
and incoherent errors and to quantify the reduction in both of them due to 
using optimal control pulses and accounting for the transfer function in an 
electron spin resonance system. We also 
prove that purity benchmarking provides bounds on the optimal fidelity and 
diamond norm that can be achieved by correcting the coherent errors through
improving calibration.
\end{abstract}

\maketitle

A key obstacle to realizing scalable quantum information processing (QIP) is 
implementing quantum gates sufficiently precisely so that errors can be detected and 
corrected~\cite{knill1997theory, knill1998resilient, 
preskill1998reliable, knill2005quantum, aliferis2007accuracy, 
gottesman1997stabilizer}. This requires both the 
intrinsic noise and the noise in the control to be characterized. The combined noise can be completely characterized using either quantum
process tomography (QPT) \cite{QPT1,QPT2} or gate set tomography (GST) 
\cite{GST1, GST2}. However, these methods are time-consuming and scale 
exponentially in the number of qubits.

Instead of completely characterizing a system, we can efficiently 
quantify how noisy the 
experimental operations are. The most prominent method along 
these 
lines is randomized benchmarking (RB)~\cite{emersonRBM, PhysRevA.77.012307, 	
dankert2009exact, PhysRevLettRobustRB, PhysRevARobustRB, RBwithConfi}, which 
gives an efficient estimate of the benchmarking error per gate (BEPG) defined as
\begin{align}
\epsilon(\mathcal{E}) = 1-F=1 - \int {\rm d}\psi\,
\bra{\psi}\mathcal{E}(\ket{\psi}\!\bra{\psi})\ket{\psi},
\end{align}
where $\mathcal{E}$ is the noise channel and the integral (the 
channel fidelity $F$) is over all pure states $|\psi\rangle$ according to the 
Haar measure. However, the BEPG is, by construction, insensitive to 
many of the particular details of the noise mechanism. As errors due to 
different noise mechanisms can be corrected in different ways and have 
different impacts on QIP, understanding the noise characteristics in quantum 
systems is of critical importance. 

Noise characteristics can be broadly grouped as either coherent 
(unitary) or incoherent (statistical). Coherent noise is usually due to
systematic control errors in, for example, imperfect rotation angles or axes 
\cite{CPMGPRA,phaseest}, which may be easier to reduce than 
incoherent noise such as $T_1$ and $T_2$ processes. The BEPG for coherent 
noise accumulates quadratically with the number of gates whereas incoherent noise accumulates linearly. 
Furthermore, coherent and incoherent noises with the same BEPG may lead to 
dramatically different thresholds as quantified by the worst-case error 
per gate (WEPG), also known as the diamond distance,~\cite{Sanders2015}
\begin{align}
\epsilon_\diamond = \tfrac{1}{2}\max_\psi \|[\mathcal{E}-\mathcal{I}]\otimes 
\mathcal{I}(\psi)\|_1,
\end{align}
where $\|A\|_1 = \Tr\sqrt{A^\dagger A}$ and $\mathcal{I}$ is the identity 
channel acting on an ancillary system of the same size to account for the 
effect of the noise on entangled inputs. Therefore, identifying whether the 
noise is primarily coherent or incoherent is essential for determining an 
appropriate error threshold when evaluating a physical system and for 
determining whether experimental effort should prioritize improving 
calibration or suppressing incoherent error processes.

Several approaches have been developed to provide 
more information about the noise than just the BEPG while retaining the 
advantages of RB~\cite{RBloss, RBPhasenoise, coherence, coherence1}. In 
particular, purity benchmarking (PB)~\cite{coherence} enables the 
quantification of the coherence of a noise process without assuming a specific 
noise model, which can be used to obtain an improved estimate of the 
WEPG~\cite{confidence,Kueng2015}, whereas the method of Ref. 
\cite{coherence1} detects additive coherent errors under specific 
assumptions about the noise model. 

In this paper, we show that PB 
can be used to quantify the best achievable BEPG and WEPG under optimal 
control for single-qubit systems. We then test PB in a specific 
modality, namely, a solid-state electron spin resonance (ESR) system. Bulk ESR samples consist of an ensemble of (nearly) identical spins, which can mimic the behaviour of a fixed number of qubits depending on the 
structure of the solid and the species of the spins. ESR provides one path to scalable QIP using techniques 
such as algorithmic cooling and distributed node quantum information 
processing~\cite{Borneman2012}, which are viable because electron spins have 
larger thermal polarization and faster relaxation rates than nuclear spins, and 
hyperfine-coupled nuclear spins can also be efficiently controlled using ESR 
techniques \cite{hodges2008universal,zhang2011coherent,HBAC_daniel}. The 
quantum control techniques developed in QIP are also very useful for modern ESR 
spectroscopy \cite{pulsedESR,Shapedoptimalcontrol}. Achieving high fidelity quantum control 
in ESR is challenging due to the limited bandwidth of a conventional microwave 
resonator. In this work, RB and PB protocols are used to assess the control 
accuracy of an ensemble single-qubit system. We demonstrate the reduction in both 
the coherent and incoherent errors obtained by first using the transfer 
function of the microwave control system to correct numerically-derived optimal 
control (OC) pulses \cite{khaneja2005optimal} and then using a spin-packet selection technique to 
effectively reduce the inhomogeneous spectral broadening \cite{ESRRB}. The 
lowest values we obtained for BEPG ($\epsilon$) and the incoherent error 
($\epsilon_{\rm in}$, defined below) for Clifford gates are $6.3\times 10^{-3}$ 
and $5.4\times 10^{-3}$, respectively.

\textit{The incoherent error per gate---}The primary characteristic of a 
coherent noise process is that it can be corrected by directly reversing the 
unitary process with perfect control. We therefore define the incoherent error 
per gate (IEPG) of a noise channel $\mathcal{E}$ to be the optimal BEPG that 
can be achieved by correcting $\mathcal{E}$ with perfect unitary operations, 
that is,
\begin{align}\label{def:incoherent_error}
\epsilon_{\rm in}(\mathcal{E}) = \min_{\mathcal{U},\mathcal{V}}
\epsilon(\mathcal{U}\circ\mathcal{E}\circ\mathcal{V})
\end{align}
for any unitary operations $\mathcal{U}$ and $\mathcal{V}$. For a general $d$-dimensional system, the incoherent error
satisfies
\begin{align}
\epsilon(\mathcal{E}) \geq \epsilon_{\rm in}(\mathcal{E}) \geq
\frac{d-1}{d}\left[1-\sqrt{u(\mathcal{E})}\right],\label{nonEq}
\end{align}
where the unitarity is \cite{coherence}
\begin{align}
u(\mathcal{E})=\frac{d}{d-1}\int \mathrm{d}\psi 
\Tr[\mathcal{E}(|\psi\rangle\!\langle\psi|-\tfrac{1}{d}\one_d)]^2.
\end{align}

We now show that the lower bound on the incoherent error in Eq. \eqref{nonEq} 
is saturated to $O[\epsilon_{\rm in}(\mathcal{E})^2]$ in the 
single-qubit case. Let $\boldsymbol{\mathcal{E}}_{j,k} = \Tr [\sigma_j^\dagger 
\mathcal{E}(\sigma_k)]/2$ be the process matrix of $\mathcal{E}$, where 
$\{\sigma_0,\sigma_1,\sigma_2,\sigma_3\} = 
\{\one_2,\sigma_x,\sigma_y,\sigma_z\}$. The process matrix of any completely-positive and 
trace-preserving (CPTP) noise channel can be written in block form as
\begin{align}
\boldsymbol{\mathcal{E}}= \left( \begin{array}{cc}
	1 & \mathbf{0} \\
	\boldsymbol{\mathcal{E}}_n & \boldsymbol{\mathcal{E}}_u  \end{array}
\right)\label{define1}.
\end{align}
The unitarity and BEPG of $\mathcal{E}$ are
\begin{align}
u(\mathcal{E})
&=\frac{1}{3}\Tr\boldsymbol{\mathcal{E}}_u^\dagger\boldsymbol{\mathcal{E}}_u
 \notag\\
\epsilon(\mathcal{E}) &=
\frac{1}{6}\Tr\left(\one_3-\boldsymbol{\mathcal{E}}_u\right).\label{define2}
\end{align}
Any single-qubit noise channel can be corrected to another channel 
$\mathcal{E}'$ such that $\boldsymbol{\mathcal{E}}'_u = \Sigma$ and 
$\boldsymbol{\mathcal{E}}'_n = (0,0,\lambda)^T$ for some $\lambda$ and 
some real diagonal $\Sigma$ by applying suitable (perfect) unitary 
operators~\cite{King2001}, which leaves the unitarity unchanged, that is, 
$u(\mathcal{E})=u(\mathcal{E}') = \Tr\Sigma^2$. By Von Neumann's trace 
inequality,
\begin{align}
\epsilon(\mathcal{E}')\leq
\epsilon(\mathcal{U}\circ\mathcal{E}\circ\mathcal{V})\label{S3}
\end{align}
for any unitary operations $\mathcal{U}$ and $\mathcal{V}$, so 
$\epsilon(\mathcal{E}') = \epsilon_{\rm in}(\mathcal{E})$. Writing $\Sigma 
=\one_3  - \epsilon(\mathcal{E}')\delta$ where $\delta$ is nonnegative 
for any CPTP map~\cite{Ruskai2002} and $\Tr\delta =6$ from Eq.~\eqref{define2}, 
we have
\begin{align}
u(\mathcal{E}) &= 1 -\frac{2\epsilon(\mathcal{E}')}{3}\Tr\delta +
\frac{\epsilon(\mathcal{E}')^2}{3}\Tr\delta^2 \notag\\
&= 1 - 4\epsilon(\mathcal{E}') + (4+c)\epsilon(\mathcal{E}')^2,\label{S4}
\end{align}
for a single qubit. The minimum and maximum values of $c$ subject to 
$\epsilon(\mathcal{E}')\leq 1/3$, $\Tr\delta =6$, and the CPTP 
constraints~\cite{Ruskai2002,RBwithConfi} are 0 and 2, attained when $\delta = 
2\one_3$ and $\delta_{1,1}=\delta_{2,2}=3$ respectively. Therefore the 
incoherent error for a single qubit satisfies
\begin{align}\label{eq:incoherent_error}
\epsilon_{\rm in}(\mathcal{E}) 
= \epsilon(\mathcal{E}')
= \frac{1}{2}\left(1-\sqrt{u(\mathcal{E})}\right),
\end{align}
to within $\epsilon_{\rm in}(\mathcal{E})^2/2$ as claimed. 

Now we consider the part of error that is removed by the optimal unitary 
corrections. With
$\mathcal{E} = \mathcal{U}\circ\mathcal{E}'\circ\mathcal{V}$ and $\mathcal{W} = 
\mathcal{V}\circ\mathcal{U}$, from Eq. (\ref{define2}) the BEPG of $\mathcal{E}$ is
\begin{align}
\epsilon(\mathcal{E}) 
&= \frac{1}{6}\Tr(\one_3 - \Sigma) + 
\frac{1}{6}\Tr(\one_3-\boldsymbol{\mathcal{W}}_u) \notag\\
&\quad  - \frac{1}{6}\Tr(\one_3-\boldsymbol{\mathcal{W}}_u) 
(\one_3-\Sigma) \notag\\
&= \epsilon_{\rm in}(\mathcal{E}) + \epsilon(\mathcal{W}) + O[\epsilon_{\rm 
in}(\mathcal{E})\epsilon(\mathcal{W})]
\end{align}
where the order of the higher-order term comes from $\Sigma$ being diagonal and 
the diagonal elements of a generic CPTP map $\mathcal{M}$ being
$1-O(\epsilon(\mathcal{M}))$~\cite{RBwithConfi}. We can regard $\mathcal{U}$ 
and $\mathcal{V}$ as coherent errors and so the BEPG of the (composite) 
coherent error is
\begin{align}
\epsilon_{\rm coh}(\mathcal{E}) = \epsilon(\mathcal{W}) = \epsilon(\mathcal{E}) 
- \epsilon_{\rm in}(\mathcal{E}) + O(\epsilon_{\rm 
	in}(\mathcal{E})\epsilon(\mathcal{W})),
\end{align}
which is also equivalent [to $O(\epsilon_{\rm in}(\mathcal{E}) 
\epsilon(\mathcal{W}))$] to the BEPG removed by the optimal unitary 
corrections.

The IEPG also provides an improved bound on the optimal WEPG 
$\epsilon_{\diamond,\rm opt}$ that can be achieved by applying unitary 
corrections. Let $\mathcal{E}'_u$ be the unital part of $\mathcal{E}'$, 
that is, the channel such that $\mathcal{E}'(A) = \mathcal{E}'_u(A) + 
\lambda \sigma_z \Tr A$ for all $A\in\mathbb{C}^{2\times 2}$. We then have
\begin{align}
\mathcal{E}'\otimes \mathcal{I}(\rho) = \mathcal{E}'_u\otimes 
\mathcal{I}(\rho) + \lambda \sigma_z\otimes \Tr_1 \rho
\end{align}
where $\Tr_1\rho$ is the partial trace over the first system.
By the triangle inequality and submultiplicativity of the diamond norm,
\begin{align}
\epsilon_\diamond(\mathcal{E}') &\leq \epsilon_\diamond(\mathcal{E}'_u) + 
|\lambda|\max_\psi\|\Tr_1 \psi\|_1 \notag\\
&\leq \epsilon_\diamond(\mathcal{E}'_u) + \sqrt{2}|\lambda|,
\end{align}
where the maximization is achieved by any maximally entangled state. As 
$\mathcal{E}'_u$ is a Pauli channel~\cite{RBT1, PhysRevARobustRB}, 
$|\lambda|\leq 3\epsilon(\mathcal{E}')$ and with a lower 
bound on the WEPG in terms of the BEPG~\cite{RBwithConfi}, we have
\begin{align}
\epsilon_{\diamond,\rm opt}(\mathcal{E}) = \epsilon_\diamond(\mathcal{E}') \in 
[\tfrac{3}{2}\epsilon_{\rm in}(\mathcal{E}),(\tfrac{3}{2} + 
3\sqrt{2})\epsilon_{\rm in}(\mathcal{E})].
\end{align}
Both these constraints are linear in $\epsilon_{\rm in}(\mathcal{E})$ and so 
give reasonable estimates as $\epsilon_{\rm in}(\mathcal{E})$ decreases 
compared to the gap between the optimal scalings for 
the lower and upper bounds in terms of $\epsilon(\mathcal{E})$ alone, which 
diverge by orders of magnitude as $\epsilon(\mathcal{E})$ 
decreases~\cite{Sanders2015}. 

\textit{Experimental Implementation---}Our X-band pulsed ESR spectrometer
was custom-built for QIP experiments and includes arbitrary waveform generation 
and a loop-gap resonator for sub-millimeter sized samples that allows for 
relatively broadband control \cite{ESRRB}. For an ensemble single-qubit system, 
we use a sample of gamma-irradiated fused quartz, a paramagnetic sample in powder form
where the primary defect is a spin-1/2 unpaired electron
at an oxygen vacancy \cite{quartz}, with $T_1\sim 160$ $\mu$s, 
$T_2\sim 30$ $\mu$s, and $T_2^*\sim 80$ ns. %All gates are implemented using OC pulses in order to evaluate the quality of microwave control with arbitrarily shaped waveforms. %Such pulses are required for controlling nearby nuclear 
%spins via 
%the anisotropic hyperfine interaction 
%\cite{hodges2008universal,zhang2011coherent}.

A pulse generated with an initial waveform $W(f)$ in the
frequency-domain representation will be distorted to a new waveform
$W'(f)$ seen by the spins due to the system's transfer function $\mathcal{T}$, which is the frequency-domain 
representation of the impulse response of the system 
\cite{Transfer1,Transfer2}, so that
$W'=\mathcal{T}\cdot W$ where $\cdot$ denotes the
point-wise product of $\mathcal{T}$ and $W$. The transfer function 
includes contributions from the resonator's transfer function and other 
imperfections in the pulse generation and transmission. One method to correct
$W'(f)$ is to distort the initial waveform to be
$\mathcal{T}^{-1}\cdot W$. The accuracy of
this method is limited by the accuracy with which $\mathcal{T}$ can be
determined. We measure $\mathcal{T}$ by detecting Rabi oscillations of the 
electron spins as a function of the microwave frequency \cite{supple}. This 
measured transfer function, denoted by $\mathcal{T}_{\rm meas}$, is then used 
to modify the input OC pulse so that the distorted pulse seen by the spins will 
approximate the desired waveform.

We use three OC pulses: $\pi/2$ and $\pi$ rotations (denoted by X90, X180, Y90 
and Y180 for rotations around the $x$- and $y$-axes respectively) and an 
identity operation (denoted by $\mathcal{I}$). The pulses are each 150 ns long and designed to be 
robust to distributions of Larmor frequency and microwave ($B_1$) field that 
closely mimic the measured properties of the combined system of our sample and 
resonator \cite{supple}. The design 
fidelity of each pulse exceeds 99.7\% when averaged over these distributions  
\cite{supple}. The experimental results span three
different conditions for implementing the OC pulses: (1) not taking the 
system transfer function into account, \textit{i.e.}, assuming $\mathcal{T}=1$ 
for all frequencies, (2) modifying the input pulses using $\mathcal{T}_{\rm 
meas}$, and (3) the same as (2) but also implementing a spin-packet selection 
(SEL) state preparation sequence \cite{ESRRB} which effectively increases 
$T_2^*$ by a factor of 2.

We implement the 24 elements of the Clifford group as $\mathbf{G}=\mathcal{S}\mathcal{P}\mathcal{Z}$ where $\mathcal{S}\in 
\{\rm \mathcal{I},X90,Y90$\}, $\mathcal{P}\in \{\rm \mathcal{I},Y180\}$,
and $\mathcal{Z}\in \{\rm \mathcal{I},Z90,Z180,Z270\}$. $\mathcal{S}$ and 
$\mathcal{P}$ are implemented using the numerically derived $\mathcal{I}$, $\pi/2$ and $\pi$ pulses and altering the phase as needed to achieve $x$- and $y$-axis 
rotations. The operations in $\mathcal{Z}$ are implemented virtually by 
changing the reference frame \cite{colmbm}. The initial state in all 
experiments is represented by the deviation density matrix $\sigma_z$.

\begin{figure}[h!]
%%\centering
\includegraphics[width=0.48\textwidth]{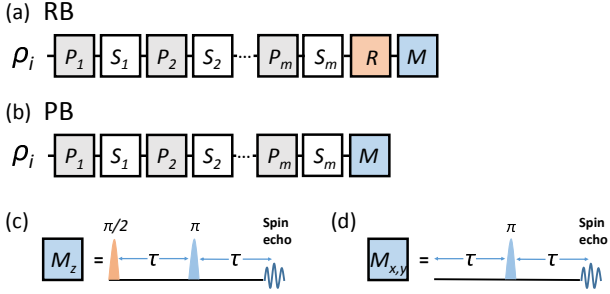}
\caption{(Color online) Quantum circuits for (a) RB and (b)
PB. The initial state $\rho_i$ is $\sigma_z$ and the 
measurements $M$ are spin echo detection sequences for measuring 
$\langle\sigma_z\rangle$ for RB and $\langle\sigma_{x,y,z}\rangle$ for PB. $R$ 
in (a) is the recovery gate that returns the state to $\pm\sigma_z$. 
A total of 150 random sequences with $S_j\in$ \{$\mathcal{I}$,~X90,~Y90\} and $P_j\in$ 
\{$\mathcal{I}$,~X180\} (and virtual $z$-axis rotations) are applied for each 
sequence length $m$ for RB and PB.
(c) and (d) are the spin echo detection sequences for measuring $\langle 
\sigma_z\rangle$ and $\langle \sigma_{x,y}\rangle$, respectively. The $\pi/2$ 
and $\pi$ pulses are 35 ns Gaussian pulses around the \emph{y}-axis, and 
$\tau$=700 ns represents a delay.}% The sequences are composed such that the Clifford gates are uniformly random.  The $\pi/2$ pulse rotates the spin magnetization along $z$-axis to $x$-axis. The $\pi$ pulse acts as the refocusing pulse. The spin echo appears at time $\tau$ after the refocusing pulse and the spin magnetization in the $x$-$y$ plane is detected inductively.}
\label{circuit}
\end{figure}

We can estimate the BEPG and IEPG averaged over the set of operations 
$\mathbf{G}$ via RB and PB as follows~\cite{PhysRevLettRobustRB,coherence} (see the quantum circuits in Fig. \ref{circuit}). (1) 
Prepare the state $\sigma_z$. (2) Apply a sequence of $m$ uniformly-random 
operations from $\mathbf{G}$, which maps $\sigma_z$ to $\rho_j$. (2.1) For RB, 
apply a recovery gate $R\in\mathbf{G}$ that maps $\rho_j$ back to 
$\pm\sigma_z$. When the final state is $-\sigma_z$ we change the sign to be 
positive in post-processing, that is, implement a virtual X gate.
(3) For RB, estimate the expectation value $\langle\sigma_z\rangle$. For PB, 
estimate the purity
\begin{align}\label{eq:purity}
P = \langle \sigma_x\rangle^2+\langle \sigma_y\rangle^2+\langle
\sigma_z\rangle^2
\end{align}
of the final state $\rho_j$. Averaging over random sequences of length $m$ and
fitting to
\begin{align}
\langle \sigma_z\rangle &=A_z + B(1-2\epsilon)^m \notag\\
\langle P\rangle &=A'+B'u^{m-1}\label{model}
\end{align}
for RB and PB, respectively, under trace-preserving noise, allows $\epsilon$ and 
the unitarity $u$ (and hence $\epsilon_{\rm in}$ via 
Eq.~\eqref{eq:incoherent_error}) to be estimated where the constants absorb 
the state preparation and measurement (SPAM) errors and the non-unitality of the noise. In particular, $A' = 
\sum_M A_M^2$ ($M\in\{\sigma_x,\sigma_y,\sigma_z\}$) with
\begin{align}\label{eq:constant}
A_M = \Tr M\mathcal{E}(\tfrac{1}{2}\one_2) = \frac{1}{24}\sum_G \Tr 
M\mathcal{E}(G\rho G^\dagger), 
\end{align}
where the summation is over the single-qubit Clifford 
group and the equality follows from the fact that the Clifford group is a 
unitary 2-design and hence is also a unitary 1-design~\cite{dankert2009exact}. 
We can therefore estimate both constant off-sets by performing a single 
Clifford gate, measuring the expectation values of $\langle\sigma_z\rangle$, 
$\langle\sigma_x\rangle$, and $\langle\sigma_y\rangle$ and averaging over all 
Clifford gates. The expectation values 
are measured by the corresponding spin echo detection sequences in Figs. 
\ref{circuit}(c) and (d). We sample 150 random sequences for each 
sequence length $m$ of RB and PB independently.

\begin{figure*}[t!]
	%%\centering
\begin{minipage}{.46\linewidth}
	\includegraphics[width=0.99\linewidth]{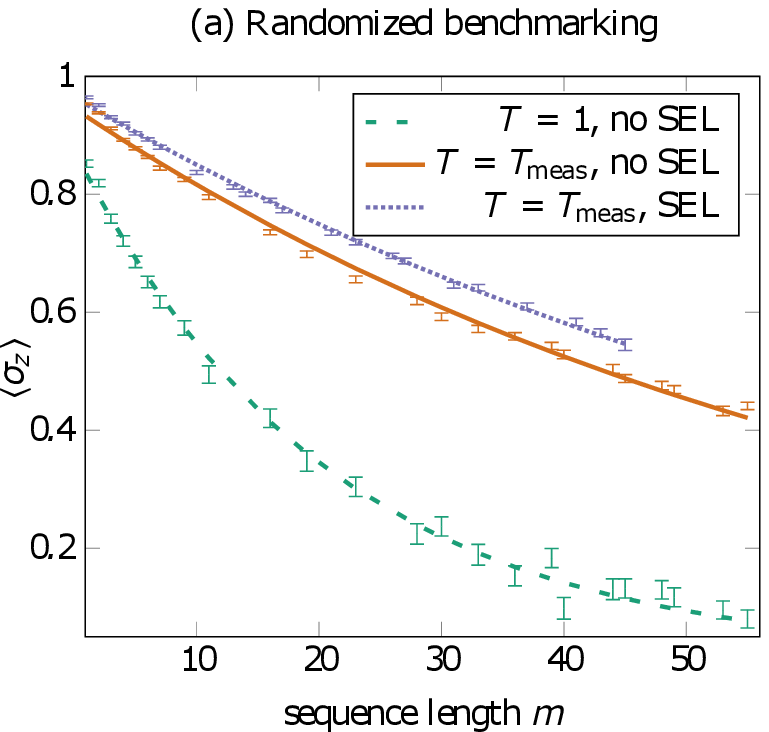}
\end{minipage}
\begin{minipage}{.46\linewidth}
	\includegraphics[width=0.99\linewidth]{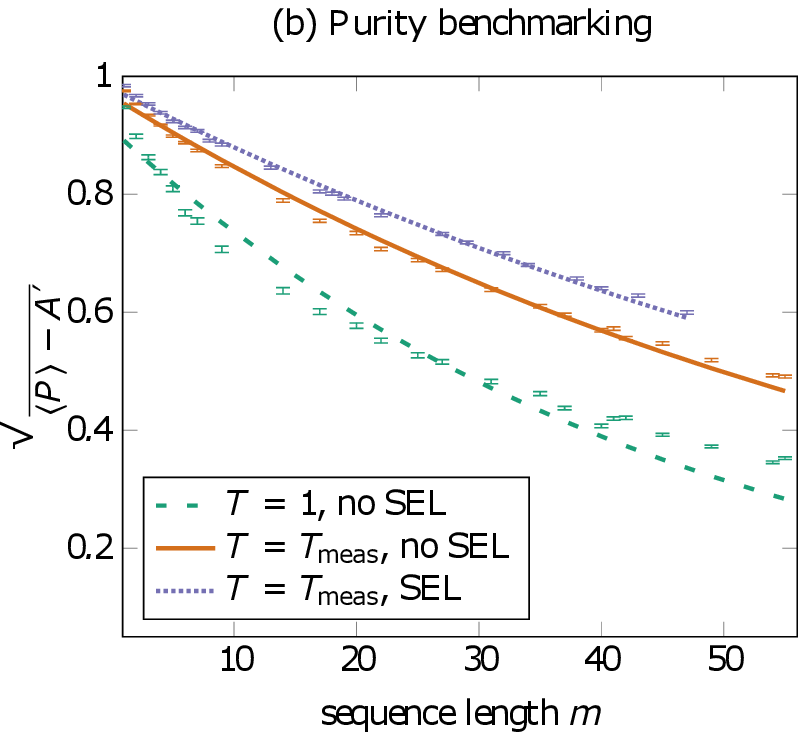}
\end{minipage}
	\caption[]{(Color online) Experimental (a) RB and (b) PB results. Each 
	experimental data point of $\langle\sigma_z\rangle$ and $\langle  P\rangle$ 
	is an average over 150 random sequences of $m$ Clifford gates where $P$ is 
	defined in Eq. \eqref{eq:purity}, and the error bars indicate the standard 
	error of the mean. The lines are least-squares fits to  $\langle 
	\sigma_z\rangle = B(1-2\epsilon)^m+A_z$ and $\langle P \rangle = 
	B'u^{m-1}+A'=B'(1-2\epsilon_{\rm in})^{2(m-1)}+A'$, respectively. $A_z$ are 
	0.0156$\pm$0.0005, 0.0009$\pm$0.0010 and -0.0004$\pm$0.0013, and $A'$ are 0.0004$\pm$0.0001, 0.0005$\pm$0.0001 and 0.0001$\pm$0.0001, for the three cases `$\mathcal{T}=1$, no SEL' (green dashed 
	line), `$\mathcal{T}=\mathcal{T}_{\rm meas}$, no SEL' (orange solid line), and `$\mathcal{T}=\mathcal{T}_{\rm meas}$, SEL' (purple dotted line), respectively. $A_z$ and $A'$ are 
	estimated using Eq. (\ref{eq:constant}). In (b), $\sqrt{\langle  
	P\rangle-A'}$ is plotted instead of $\langle  
	P\rangle$ to show that $\sqrt{\langle  P\rangle-A'}$ has a slower decay than 
	$\langle\sigma_z\rangle$, indicating 
	$\epsilon_{\rm in}<\epsilon$. The $\epsilon$  and 
	$\epsilon_{\rm in}$ values are given in Table~\ref{tab:parameters}.  Because of the limitation of the pulsed traveling wave tube amplifier, the largest $m$ are 55 and 47 in the cases without and with SEL sequences, respectively. All $m$ are chosen randomly and independently for RB and PB sequences.}
	\label{regular}
\end{figure*}

\textit{Results and Discussion---}The results of the RB and PB experiments are
presented in Fig.~\ref{regular}, with the corresponding estimates for the
BEPG, IEPG, coherent error rate and optimal WEPG listed in Table~\ref{tab:parameters}.

\begin{table}[h!]
\normalsize
\centering
\begin{tabular}{l C{24mm} C{24mm} C{24mm}}
	\hline\hline
	& $\mathcal{T}=1$ & $\mathcal{T}=\mathcal{T}_{\rm meas}$ & 
	$\mathcal{T}=\mathcal{T}_{\rm meas}$ \\
	& no SEL & no SEL & SEL \\\hline
$\epsilon$ & 0.0234(11) & 0.0073(2) &0.0063(2)  \\
$\epsilon_{\rm in}$ & 0.0105(10) & 0.0066(2) &0.0054(2) \\
$\epsilon_{\rm coh}$ & 0.0129(21) & 0.0007(4) & 0.0009(4) \\
$\epsilon_{\diamond,\rm opt}$ & 0.040(26) & 0.024(15) & 0.020(12) \\
	\hline\hline
\end{tabular}
\caption[Results]{Estimates of the BEPG $\epsilon$, IEPG $\epsilon_{\rm in}$, 
coherent error rate $\epsilon_{\rm coh}$ and optimal WEPG under perfect 
calibration $\epsilon_{\diamond,\rm opt}$ per Clifford gate. Gates are realized with OC pulses that 
assume a flat transfer function ($\mathcal{T}=1$) or are distorted 
based on the measured transfer function ($\mathcal{T} = \mathcal{T}_{\rm 
meas}$), and with or without spin packet selection (SEL) sequences 
respectively. Note that the values listed here are obtained by fitting the RB and PB data to a single-exponential decay, whereas the actual decays are non-exponential, especially noticeable in the $\mathcal{T}=1$ case. Thus, the estimated gate errors given here are effectively averaged over the non-Markovian noise (see main text).}% The error bars indicate the 95\% confidence interval.}%The error bars come from the fitting to the models in Eq. \ref{model} and indicate the 95\% confidence interval.}  
\label{tab:parameters}
\end{table}

Pulse distortion due to the system transfer function is significant, as the 
transfer function bandwidth of $\sim$100 MHz is comparable to the pulse 
excitation bandwidth. The improvement between results from the unmodified OC 
pulses ($\mathcal{T}=1$) and those modified by taking into account $\mathcal{T}_{\rm 
meas}$ in Table~\ref{tab:parameters} demonstrates substantial reduction in $\epsilon_{\rm coh}$ from $\sim 10^{-2}$ to $\sim 
10^{-3}$.
The $\epsilon_{\rm in}$ is also reduced by approximately a factor of two, from $\sim 1.0\times 10^{-2}$ 
to 
$\sim 0.5\times 10^{-2}$. This shows that the pulse distortion is 
non-negligible, and causes both coherent and incoherent errors. The larger 
incoherent error for the unmodified OC pulses is largely due to the OC pulses 
losing their engineered robustness to Larmor frequency and $B_1$ inhomogeneities 
when assuming $\mathcal{T}=1$.

%both the average coherent error ($\epsilon-\epsilon_{\rm in}$) and the incoherent error. This shows that the pulse distortion is non-negligible, and causes error that contains both coherent and incoherent parts. The larger incoherent error for the unmodified OC pulses is largely due to the OC pulses losing their robustness to $B_0$ and $B_1$ field inhomogeneities when assuming $\mathcal{T}=1$. %incoherent part is 
%largely due to the OC pulses losing their robustness to $B_0$ and $B_1$ field 
%inhomogeneities.  seen when neglecting the transfer function 

Although the decay rates of both the RB and PB experimental results are 
substantially reduced by using  $\mathcal{T}_{\rm meas}$ to improve the OC 
pulses, the decays seem to
deviate from a single exponential decay (i.e., see the oscillating deviations of the orange data points from the 
orange solid lines in Fig.~\ref{regular}), implying the existence of 
non-Markovian noise. In our system, the Larmor frequency 
distribution for different spin-packets ($T_2^*$ effect) results in a 
significant non-Markovian effect \cite{ESRRB,nonexponential}. The benchmarking pulse sequences 
act like filters, in that the spectral line-width of the part of the 
spin-packet that contributes to the signal decreases with the number of gates. 
This means the effective $T_2^*$ lifetime is not constant but increases with the number of gates that are implemented. Therefore, the error rates estimated using the single-exponential decay model are the averaged values over this non-constant noise. Lindblad numerical 
simulations (where the $T_2^*$ process is simulated by averaging over multiple 
simulations 
with different Larmor frequencies) give non-exponential decays for RB and PB 
\cite{supple}, agreeing with our experimental results. To reduce 
the non-Markovianity due to $T_2^*$, we implement SEL sequences before each 
of the benchmarking sequences, which selects a narrower line-width so 
the benchmarking experiments have a longer $T_2^*$ ($\sim$ 160 ns) to begin 
with \cite{supple}. After incorporating the SEL sequences, the experimentally observed decays fit to a single exponential better (see the purple data points and purple dotted lines in Fig.~\ref{regular}).  The Lindblad simulation results with the longer 
$T_2^*$ also exhibit single exponential decays up to $\sim$ 50 gates  \cite{supple}.%decays consistent with a single exponential for up to $\sim$50 gates \cite{supple}, which is consistent with our observed experimental data. which are multiple 2$\pi$ rotations sensitive to the Larmor frequency distribution.

Using the SEL sequence improves $\epsilon_{{\rm in}}$ from $(6.6\pm 0.2)\times 
10^{-3}$ to $(5.4\pm 0.2)\times 10^{-3}$, but has no statistically
significant effect on $\epsilon_{\rm 
coh}$, which is $(0.9\pm 0.4)\times 10^{-3}$  and  $(0.7\pm 0.4)\times 10^{-3}$
with and without SEL, respectively. This implies the $T_2^*$ effect mainly 
contributes to the incoherent error. In the Lindblad simulations of the 
benchmarking sequences using the extended $T_2^*$, $\epsilon_{{\rm in}}$ caused 
by  $T_1$, $T_2$, and $T_2^*$ is $3.5\times 10^{-3}$, and  $\epsilon_{{\rm 
coh}}$ caused by the imperfection in the OC pulse design is $0.5\times 10^{-3}$ 
\cite{supple}. We attribute the discrepancy between the simulated and 
experimental values of $\epsilon_{{\rm in}}$ and  $\epsilon_{{\rm 
coh}}$  to possible inaccuracy in the measured decoherence times, 
fluctuations in the control mechanisms, and imperfect knowledge of the transfer 
function.

\textit{Conclusions--} We have demonstrated how RB and PB can be used together 
to go beyond quantifying average gate fidelities by distinguishing coherent and 
incoherent contributions to the error. This allows improvements in 
calibration and engineering pulses to suppress incoherent errors to be 
implemented and diagnosed independently. Pulse distortion due to the system 
transfer function $\mathcal{T}$ is the dominant error source in our system and 
contributes greatly to the coherent part of the gate error. Our measurement of 
$\mathcal{T}$ helps improve the OC 
pulse fidelities significantly. The incoherent error is primarily due to $T_1$, $T_2$ and
$T_2^*$ processes. By effectively extending $T_2^*$ we reduce the non-Markovian 
effect and improve the control fidelity further. %Using an improved method, or combining with other methods \cite{Transfer1, Transfer2} to improve the accuracy of $\mathcal{T}_{\rm meas}$, should enable higher control fidelities.

Results from gate set tomography included in the supplemental material 
indicate that our system has substantial gate-dependent noise. The PB protocol 
has only been analyzed under the assumption of gate-independent noise. 
Simulations using the estimates from gate set tomography indicate that PB can 
distinguish between gate-dependent coherent errors that look incoherent when 
averaged over the gates and a gate-independent incoherent process, at least for 
some physically-realistic error models. However, we leave the general behavior 
of PB under gate-dependent noise as an open problem.

{\bf Acknowledgement} This research was supported by NSERC, the Canada 
Foundation for Innovation, CIFAR, the province of Ontario, Industry Canada, the Gerald Schwartz \& Heather Reisman Foundation, and 
the U.S. Army Research Office through grant W911NF-14-1-0103. We thank David Cory and Troy Borneman for providing the ESR sample and for stimulating discussions; Colm 
Ryan, Yingjie Zhang, and Jeremy Chamilliard for their contributions to the 
spectrometer; Ming Lyu and Ian Hincks for the discussions on the transfer function measurement method; Robabeh Rahimi for help with preparing the ESR sample; Roberto Romero and Hiruy Haile for help with machining; Hemant Katiyar for help with the simulations.

%\bibliography{UDA}

\vskip 12pt

\clearpage

\onecolumngrid

\begin{widetext}
\center
{\bf Supplementary Material: Estimating the coherence of noise in quantum control of a solid-state qubit}
\medskip
\bigskip
\end{widetext}

\twocolumngrid
\appendix

\section{Determination of the system transfer function}
%\label{sec:intro}
In order to characterize our transfer function, we measure Rabi oscillations of the spin signal under microwave pulses across a set of offset frequencies. By numerically fitting the measured oscillations to a theoretical model, the amplitude and phase of the transfer function are obtained. 

First, we consider the Hamiltonian of the single-qubit system in the lab frame:
\begin{align}
H_{\textrm{lab}}=H_{0}+H_{\textrm{mw}}=\omega_0\frac{\sigma_z}{2}+\Omega\frac{\sigma_x}{2}\cos(\omega (t-t_0)+\psi),\tag{S1}\label{Hamiltonian}
\end{align}
where $H_{0}$ is the Zeeman interaction between the electron spin and the static magnetic field and $H_{\textrm{mw}}$ is the interaction between the electron spin and the microwave field. Using the rotating-wave approximation (RWA), the Hamiltonian in the rotating frame with frequency $\omega$ has the following form:
\begin{align}
H_{\textrm{RWA}}=-\Delta\frac{\sigma_z}{2}+\Omega(\frac{\sigma_x}{2}\cos(\psi)+\frac{\sigma_y}{2}\sin(\psi)),\tag{S2} \label{rotateH}
\end{align}
where $\Delta=\omega-\omega_0$, $\Omega$, and $\psi$ are the offset frequency, amplitude, and phase of the microwave pulse, respectively. Here, we consider a constant $\Omega$ for simplicity which means the microwave pulse has a single frequency component $\omega$. $\Omega\cos(\psi)$ and $\Omega\sin(\psi)$ are always referred to as in-phase part (denoted as $\mathcal{W}_{0}$) and quadrature part (denoted as $\mathcal{W}_{90}$). This allows the waveform to be conveniently written in a complex form of $\mathcal{W}_{0}+i\mathcal{W}_{90}=\Omega e^{i\psi}$. 
Distortion caused by the finite bandwidth of the resonator and imperfections in microwave generation and transmission makes $\Omega e^{i\psi}$ different from the intended waveform $\Omega_0 e^{i\psi_0}$. The ratio $\Omega e^{i\psi}/(\Omega_0 e^{i\psi_0})$ is the value of the transfer function $\mathcal{T}$ at offset frequency $\Delta$. Therefore, $\mathcal{T}$  can be obtained by measuring $\Omega e^{i\psi}/(\Omega_0 e^{i\psi_0})$ at different offset frequencies.

\begin{figure}[h!]
%%\centering
\includegraphics[width=0.25\textwidth]{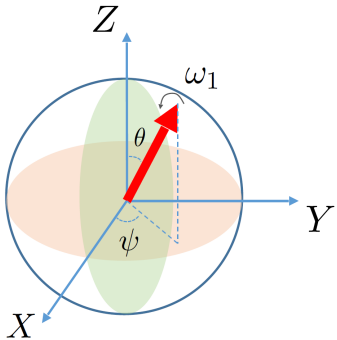}
\caption[]{(Color online) The red arrow is the rotation axis ($\sin(\theta)\cos(\psi)$, $\sin(\theta)\sin(\psi)$, $\cos(\theta)$) of the spin polarization under the Hamiltonian in Eq. (\ref{rotateH}).}
\label{sphere}
\end{figure}

\begin{figure}[h!]
%%\centering
\includegraphics[width=0.42\textwidth]{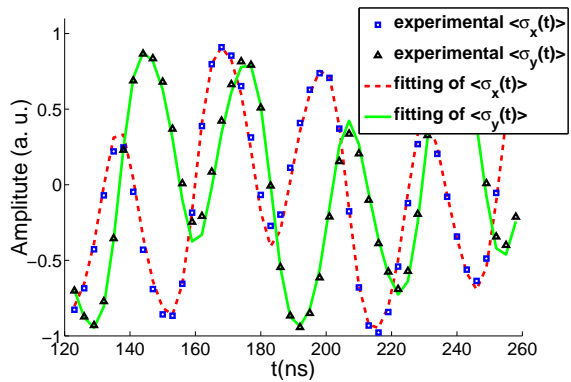}
\caption[]{(Color online) Experimental and fitted trajectories of $\langle\sigma_x(t)\rangle$ and $\langle\sigma_y(t)\rangle$ at offset frequency $\Delta=32$ MHz.}
\label{rabi}
\end{figure}

 Equation (\ref{rotateH}) shows that in the rotating frame with frequency $\omega$, the spin polarization is rotating at the frequency $\omega_1=\sqrt{\Delta^2+\Omega^2}$ around the axis $\hat{n}=(\frac{\Omega}{\omega_1}\cos(\psi),\frac{\Omega}{\omega_1}\sin(\psi),-\frac{\Delta}{\omega_1})$ (Fig. \ref{sphere}). The initial state of $\sigma_z$ evolves as:
\begin{align}
\rho(t)&=(\cos^2(\frac{\omega_1(t-t_0)}{2})+\sin^2(\frac{\omega_1(t-t_0)}{2})\cos2\theta)\sigma_z\nonumber\\\nonumber&+(\sin^2(\frac{\omega_1(t-t_0)}{2})\sin2\theta\cos\psi\nonumber\\\nonumber&+\sin(\omega_1(t-t_0))\sin\theta\sin\psi)\sigma_x\\\nonumber&+(\sin^2(\frac{\omega_1(t-t_0)}{2})\sin2\theta\sin\psi\nonumber\\\nonumber&-\sin(\omega_1(t-t_0))\sin\theta\cos\psi)\sigma_y,\nonumber\tag{S3}
\end{align}
where $\sin\theta=\frac{\Omega}{\omega_1}$ and $\cos\theta=-\frac{\Delta}{\omega_1}$.  In our experiments, it is more convenient to work in the rotating frame of frequency $\omega_0$, where the evolution of expectation values $\langle\sigma_x(t)\rangle$ and $\langle\sigma_y(t)\rangle$ can be expressed as:
\begin{align}
\langle\sigma_x(t)\rangle={\rm Tr}(\sigma_xe^{i(t-t_0)\Delta\sigma_z/2}\rho(t)e^{-i(t-t_0)\Delta\sigma_z/2})\label{sigmax}\tag{S4}\\
\langle\sigma_y(t)\rangle={\rm Tr}(\sigma_ye^{i(t-t_0)\Delta\sigma_z/2}\rho(t)e^{-i(t-t_0)\Delta\sigma_z/2}).\label{sigmay}\tag{S5}
\end{align} 
We applied constant amplitude pulses at frequency $\omega$ and fitted the experimental oscillations of $\langle\sigma_x(t)\rangle$ and $\langle\sigma_y(t)\rangle$ to Eqs. (\ref{sigmax}) and (\ref{sigmay}) (see Fig. \ref{rabi}) to extract $\psi$ and $\Omega$. By varying  $\omega$, we can obtain estimates of $\Omega e^{i\psi}/(\Omega_0 e^{i\psi_0})$ at different frequencies.  %Figure \ref{transfer} shows the estimated transfer function $\mathcal{T}$ in the offset frequency range of [-100,100] MHz.

\begin{figure}[h!]
%%\centering
\includegraphics[width=0.40\textwidth]{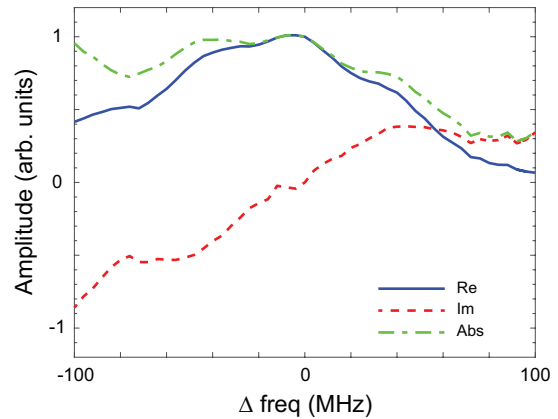}
\caption[]{(Color online) The measured transfer function, $\mathcal{T}_{\rm meas}$, as a function of offset frequency relative to the resonant frequency  $\omega_0$=9.996 GHz.}
\label{transfer}
\end{figure}

There are two issues of the method described above that need to be addressed. First, the method assumes the input microwave pulse with constant amplitude and single frequency component $\omega$. However, in real systems the pulses will have a rising and falling at the beginning and end, during which the spin experiences varying microwave field. If the rising and falling time is very short compared to the total pulse length, the effect can be neglected to a good approximation. In our system, due to the finite bandwidth of the resonator as well as limitations in the transmission components (\textit{e.g.}, IQ mixer and amplifier), the rising and falling time of a pulse is around 5$\sim$10 ns, which is non-negligible for short pulses. To reduce the effect of the rising and falling of the pulse, we used relatively long pulses (>120 ns). By fitting Rabi oscillations assuming the pulses are perfect pulses with constant amplitude, we obtained an initial guess for the transfer function (denoted as $\mathcal{T}_0$). In the next fitting iteration, we simulated the Rabi oscillations due to the pulses distorted by $\mathcal{T}_0$. In this way, we minimize the effect of imperfect pulses and obtained a refined transfer function. Another issue is that the fitting results of the phase part of the transfer function, $\psi(\omega)$, strongly depend on the choice of the starting time point $t_0$. From Eq. (\ref{Hamiltonian}), it is not difficult to prove that with a temporal shift $\delta t$ relative to $t_0$, $\psi(\omega)$ will get an additional slope which is $\delta t\omega$. Due to the distortion of the pulses, there is uncertainty in determining $t_0$. To compensate, we used the refined fitting result of the transfer function as an initial guess to modify our optimal control (OC) pulses and measured the average gate fidelity using randomized benchmarking (RB). We then performed a few iterations of feed-back control where we slightly adjusted the slope of the phase of the transfer function (and our OC pulses accordingly) until we maximized our average gate fidelity and got  the measured transfer function $\mathcal{T}_{\rm meas}$ as shown in Fig.  \ref{transfer}.  The gate set tomography (GST) results of unmodified OC  pulses, the OC pulses modified by $\mathcal{T}_{\rm meas}$, and the OC pulses modified by $|\mathcal{T}_{\rm meas}|$ are given in Table \ref{GST}. The table clearly shows both the amplitude and phase of the transfer function are important in improving the fidelities of the OC pulses.

It should be noted that $\mathcal{T}_{\rm meas}$ becomes inaccurate when the offset frequency $\Delta$ is large, as the fitting of the experimental Rabi oscillations is less sensitive to the microwave pulse amplitude and phase when $|\Omega/\Delta|$ is small. However, because the excitation bandwidths of our OC pulses are about 100 MHz or less, the imperfections of $\mathcal{T}_{\rm meas}$ for $|\Delta|$ larger than the bandwidths does not affect the controls of the OC pulses significantly. From the experimental results shown in the main text and the results listed in Table \ref{GST}, correcting the OC pulses using $\mathcal{T}_{\rm meas}$ obtained by the method described above improves the control fidelities greatly. Therefore, we conclude that our $\mathcal{T}_{\rm meas}$ is a good estimate of the system transfer function $\mathcal{T}$ within the bandwidths of the OC pulses.

%It should be mentioned that the method of measuring $\mathcal{T}$ described above assumes the input microwave pulse has only one frequency component $\omega$; however, unavoidable rising and falling time of the pulse the spin experiences varying microwave field. We placed an antenna near the sample in the resonator to monitor the rising and falling shape of the square microwave pulse \cite{ESRRB}, which helped refine the fitting results of the $\psi$  (we found the rising and falling time mainly affects $\psi$). Unfortunately, the antenna also distorts the pulse shape. To compensate, the refined antenna results were used as an initial guess followed by a few iterations of feed-back control to maximize the average gate fidelity of modeified OC pulses in RB which optimized the phase information of the transfer function. The GST results of unmodified OC  pulses, the OC pulses modified by $\mathcal{T}$ and the OC pulses modified by $|\mathcal{T}|$ are given in Table \ref{GST}. Their comparison shows both the amplitude and phase of the transfer function are important in improving OC pulses.

\section{Schematic diagram of experimental setup}
\begin{figure}[h!]
%%\centering
\includegraphics[width=0.42\textwidth]{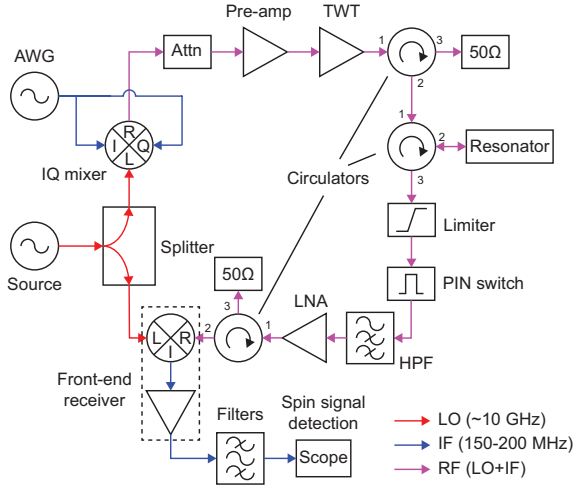}
\caption[]{(Color online) Schematic diagram of the home-built pulsed X-band ESR spectrometer. Abbreviations used in the diagram are: AWG, arbitrary waveform generator; Attn, attenuator; Pre-amp, pre-amplifier; TWT, traveling wave tube amplifier; HPF, high-pass filter; LNA, low-noise amplifier. Red arrows (LO) denote the reference microwave frequency generated by the source, blue arrows (IF) denote the frequency of the signals from the arbitrary waveform generator going to the IQ mixer for the up-conversion (spin signals have the same frequency after the down-conversion by the front-end receiver mixer), and purple arrows denote the frequency of up-converted signal.}
\label{schematic}
\end{figure}

Figure~\ref{schematic} shows the schematic diagram of the home-built pulsed X-band electron spin resonance (ESR) spectrometer used for the experiments presented in the paper. First, the signal source (Rhode $\&$ Schwarz) generates continuous-wave reference microwave signal of $\sim$10 GHz, which is split into two by the two-way splitter (Marki). For the transmission, pulses with desired amplitudes and phases are constructed by up-converting the reference using the IQ mixer (Marki) and the arbitrary waveform generator (Tektronix) which has maximum resolution of 1 ns. Typical intermediate frequency (IF) fed into the IQ ports for the up-convertion is from 150 to 200 MHz. Next the pulses enter the microwave resonator located at the center of the water-cooled electromagnet through a series of attenuator (Advanced Technical Materials) and amplifiers (pre-amplifier: MITEQ; pulsed traveling wave tube amplifier: applied system engineering) to achieve appropriate amplifications, followed by circulators (DiTom) to minimize unwanted reflections.

Spin signals are directed to the receiver which consists of the limiter (Pasternack), fast PIN didoe switch (Advanced Technical Materials), filters (Mini-circuits), low-noise amplifier (MITEQ), circulator (DiTom), and front-end receiver mixer (MITEQ). The limiter and PIN switch are implemented to prevent too much power flooding into the low-noise amplifier, and the high-pass filter (Mini-Circuits) blocks the low-frequency transient responses of the switching on and off. The circulator also protects the low-noise amplifier from unwanted reflections as well as possible leakage of the reference through the receiver mixer. Finally another filtering stage (Mini-Circuits) removes artifacts from the receiver mixer, and the spin signals, down-converted to the same IF frequency used in the up-conversion scheme, are captured by a fast digital oscilloscope (LeCroy). Further signal processing, {\it i.e.}, down-convertion to d.c., is performed by a computer. More details about the spectrometer can be found in Ref.~\cite{ESRRB}.

\section{Correction of non-linear amplitude fluctuation and phase droop of pulsed TWT amplifier}
\begin{figure}[h!]
%%\centering
\includegraphics[width=0.42\textwidth]{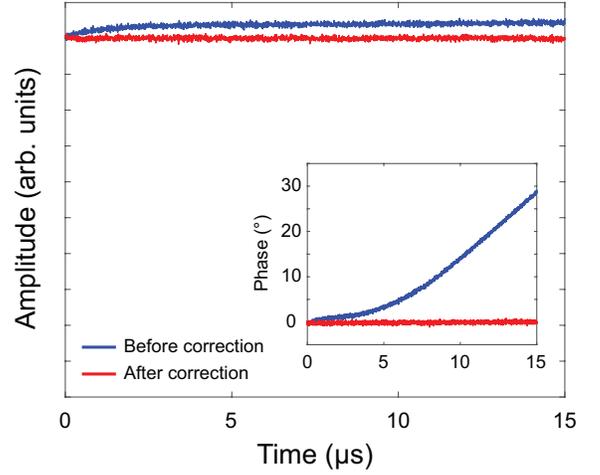}
\caption[]{(Color online) Amplitude of a long square pulse output by the TWT amplifier before and after correction measured by a directional coupler. Inset shows the phase droop. In all of the experiments, a waiting time of 2 $\mu$s after unblanking the TWT is employed to allow the output to stabilize (not shown).}
\label{twtcorrection}
\end{figure}

As we reported in Ref.~\cite{ESRRB}, the raw output from the pulsed traveling wave tube (TWT) amplifier in our setup can be non-linear (both in amplitude and phase; mainly due to heating of the amplifier) across the desired pulse sequence period. In RB and purity benchmarking (PB) experiments where we extend the sequences nearly to the maximum pulse sequence duration ($\sim$15$\mu$s) limited by the pulsed TWT amplifier, we corrected the non-linear amplitude fluctuation and phase droop in order to achieve the best control fidelities. The non-linearity was corrected using a technique similar to how the transfer function was used to pre-distort the OC pulses. We first measure the TWT amplifier's output when a long square pulse was input in order to observe the amplitude non-linearity and  phase droop as a function of time. We can feed back this information to pre-distort the input pulse sequences. After this process, the majority of non-linearity and phase droop is corrected as shown in Fig.~\ref{twtcorrection}.

\begin{table}[h!]
%\tiny
\scriptsize
\centering
\begin{tabular}{  M{1.2cm} M{1.2cm} M{1.2cm} M{1.2cm}  M{1.2cm} M{1.2cm}}
\hline \hline
\multicolumn{1}{   M{1.2cm} }{} &\multicolumn{1}{M{1.2cm} }{$F$(X90)} &\multicolumn{1}{ M{1.2cm} }{$F$(Y90)}&\multicolumn{1}{ M{1.2cm} }{$F$(X180)} &\multicolumn{1}{ M{1.2cm} }{$F$(Y180)} &\multicolumn{1}{ M{1.2cm} }{$F$($\mathcal{I}$)}\\\hline
 \multicolumn{1}{  M{1.45cm} }{$\mathcal{T}=\mathcal{T}_{\rm meas}$, SEL} &\multicolumn{1}{ M{0.75cm} }{$0.9940_{-42}^{+24}$} &\multicolumn{1}{ M{0.75cm} }{$0.9969_{-45}^{+12}$}&\multicolumn{1}{ M{0.75cm} }{$0.9926_{-43}^{+29}$} &\multicolumn{1}{ M{0.75cm} }{$0.9932_{-41}^{+25}$} &\multicolumn{1}{ M{0.75cm} }{$0.9890_{-44}^{+33}$}\\ \cline{1-6}
 \multicolumn{1}{  M{1.45cm} }{$\mathcal{T}=\mathcal{T}_{\rm meas}$, no SEL} &\multicolumn{1}{ M{0.75cm} }{$0.9914_{-26}^{+19}$} &\multicolumn{1}{ M{0.75cm} }{$0.9926_{-29}^{+23}$}&\multicolumn{1}{ M{0.75cm} }{$0.9916_{-31}^{+19}$} &\multicolumn{1}{ M{0.75cm} }{$0.9924_{-31}^{+26}$} &\multicolumn{1}{ M{0.75cm} }{$0.9838_{-26}^{+33}$}\\ \cline{1-6}
\multicolumn{1}{  M{1.45cm} }{$\mathcal{T}=1$, no SEL} &\multicolumn{1}{ M{0.75cm} }{$0.9785_{-113}^{+41}$} &\multicolumn{1}{ M{0.75cm} }{$0.9790_{-98}^{+50}$}&\multicolumn{1}{ M{0.75cm} }{$0.9796_{-130}^{+60}$} &\multicolumn{1}{ M{0.75cm} }{$0.9773_{-124}^{+48}$} &\multicolumn{1}{ M{0.75cm} }{$0.9588_{-114}^{+102}$}\\ \cline{1-6}
\multicolumn{1}{  M{1.45cm} }{$\mathcal{T}=|\mathcal{T}_{\rm meas}|$, no SEL} &\multicolumn{1}{ M{0.75cm} }{$0.9915_{-57}^{+33}$} &\multicolumn{1}{ M{0.75cm} }{$0.9912_{-50}^{+29}$}&\multicolumn{1}{ M{0.75cm} }{$0.9910_{-50}^{+36}$} &\multicolumn{1}{ M{0.75cm} }{$0.9900_{-50}^{+33}$} &\multicolumn{1}{ M{0.75cm} }{$0.9765_{-60}^{+45}$}\\\cline{1-6}
\multicolumn{1}{  M{1.45cm} }{5\% miscal, $\mathcal{T}=\mathcal{T}_{\rm meas}$, no SEL} &\multicolumn{1}{ M{0.75cm} }{$0.9880_{-36}^{+28}$} &\multicolumn{1}{ M{0.75cm} }{$0.9892_{-35}^{+21}$}&\multicolumn{1}{ M{0.75cm} }{$0.9906_{-35}^{+24}$} &\multicolumn{1}{ M{0.75cm} }{$0.9911_{-34}^{+33}$} &\multicolumn{1}{ M{0.75cm} }{$0.9811_{-34}^{+36}$}\\ \cline{1-6}
\multicolumn{1}{  M{1.45cm} }{-5\% miscal, $\mathcal{T}=\mathcal{T}_{\rm meas}$, no SEL} &\multicolumn{1}{ M{0.75cm} }{$0.9906_{-35}^{+22}$} &\multicolumn{1}{ M{0.75cm} }{$0.9924_{-31}^{+24}$}&\multicolumn{1}{ M{0.75cm} }{$0.9911_{-30}^{+26}$} &\multicolumn{1}{ M{0.75cm} }{$0.9913_{-32}^{+26}$} &\multicolumn{1}{ M{0.75cm} }{$0.9819_{-32}^{+32}$}\\  \cline{1-6} \cline{1-6}
\end{tabular}

\caption[]{Fidelity $F$ (Eq. (1)) from GST results under different experimental conditions. `$\mathcal{T}=\mathcal{T}_{\rm meas}$' and `$\mathcal{T}=|\mathcal{T}_{\rm meas}$|' denote the cases of OC pulses modified by taking into account $\mathcal{T}_{\rm meas}$ and $|\mathcal{T}_{\rm meas}|$, respectively. `$\mathcal{T}=1$' denotes the case of unmodified OC pulses. `SEL' stands for the spin-packet selection sequences. `5\% miscal' and `-5\% miscal' denote the cases when implementing OC pulses with powers that are 5\% larger and smaller than the calibrated pulse power, respectively.  The error bars are calculated using the best and worst process matrices when sampling the parameters of the process matrices within two standard deviations under the CPTP constraints.} 
\label{GST}
\end{table}

\section{Larmor frequency distribution and $\bf B_1$ field distribution}
\begin{figure}[h!]
%%\centering
\includegraphics[width=0.42\textwidth]{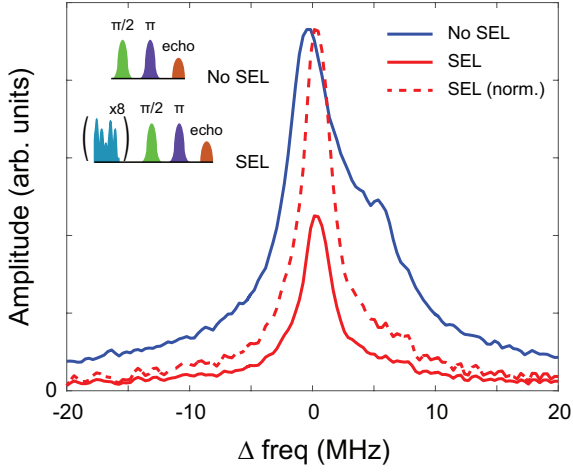}
\caption[]{(Color online) Thermal spectrum (Larmor frequency distribution profile) with and without the spin-packet selection (SEL) sequence. With the SEL sequence, the linewidth narrows by about half, but at the same time, more than half of the amplitude is lost (the dotted line which represents the spectrum with SEL sequence that is normalized to the spectrum without SEL sequence). The inset shows the pulse sequences used to measure the corresponding spectra. For SEL, a numerically derived $2\pi$ pulse that is not robust to Larmor frequency distribution is repeated eight times and each separated by an order of $T_2$ before the spin echo readout.}
\label{spectrum}
\end{figure}

\begin{figure}[h!]
\includegraphics[width=0.45\textwidth]{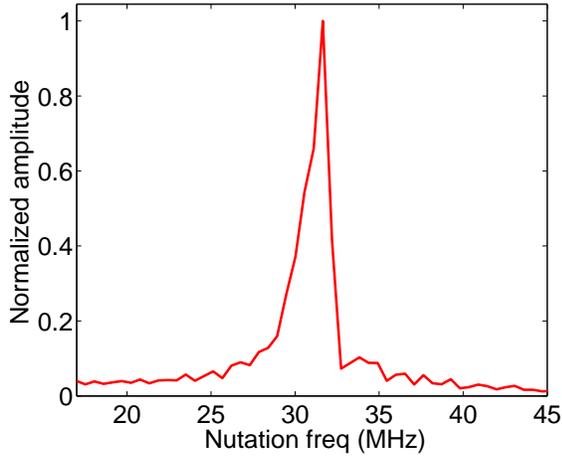}
\caption[]{(Color online) $B_1$ distribution profile obtained by taking the Fourier transform
of the experimental Rabi oscillations, and refined by comparing the experimental and simulated Rabi oscillations \cite{ESRRB}. The FWHM is
about 1.7 MHz.}
\label{B1}
\end{figure}

The Larmor frequency distribution and $B_1$ field distribution, measured by detecting the spin signals, are the properties of the combined system of the sample and the resonator (Figs. \ref{spectrum} and \ref{B1}). In particular,  the asymmetric shape of the Larmor frequency distribution is mainly due to the anisotropy of the $g$-value of the electron spins in the irradiated fused quartz sample.

\section{Optimal control pulses}
We designed three OC pulse shapes to realize $\pi/2$ and $\pi$ rotations around $x$ and  $y$-axes and an identity operation.  Different rotation axes for the $\pi/2$ and $\pi$ pulses were realized through phase shifting.  X90 and Y90 denote a $\pi/2$  pulse about the $x$- and $y$-axes, X180 and Y180 denote a $\pi$ pulse about the $x$- and $y$-axes, and $\mathcal{I}$ denotes an identity pulse. We used the gradient ascent pulse engineering algorithm \cite{khaneja2005optimal} to design OC pulses. In our system, $T_2^*<100\operatorname{ns}$ is the shortest time scale at which decoherence occurs. To make the OC pulses robust to $T_2^*$ noise, the weighted average fidelity over the Larmor frequency distribution (Fig. \ref{spectrum}) is numerically optimized. Furthermore, OC pulses are made to be robust to the inhomogeneity in the microwave field (Fig. \ref{B1}) using the same method. 

Table \ref{numerical} gives the weighted average fidelity (Eq. (1)) of each OC pulse under different numerical simulation conditions ($T_1$ and $T_2$ effects are not considered here), and it shows the ideal OC pulses should be very robust to $T_2^*$ noise and power mis-calibration. However, in the experiments, when the OC pulses arrive at the spins, they are distorted by the system transfer function. To investigate the effect of the system transfer function on the OC pulse fidelities, we use $\mathcal{T}_{\rm meas}\cdot W$ to approximately simulate the distorted OC pulses, and the fidelities are presented in Table \ref{numerical}. It shows that the coherent error of the pulses increases greatly, and the incoherent error also increases because the pulses lose part of the designed robustness to the $T_2^*$ noise and $B_1$ field distribution due to the transfer function. 

The experimental fidelities of the OC pulses are given in Table \ref{GST}. After using the measured transfer function $\mathcal{T}_{\rm meas}$ to modify the pulses to compensate the effect of the system transfer function in experiment,  despite the imperfection of $\mathcal{T}_{\rm meas}$, the pulses remain somewhat robust to these inhomogeneities. %The GST data of the GRAPE pulses with 5\% larger or smaller pulse power than the calibrated one also imply that the coherent error doesn't simply come from power miscalibration in our case.

\begin{table}[h!]
\footnotesize
\centering
\begin{tabular}{  M{1.5cm} M{1.2cm} M{1.2cm} M{1.2cm} M{1.2cm} M{1.2cm}}
\hline \hline
\multicolumn{1}{   M{1.5cm} }{} &\multicolumn{1}{M{1.2cm} }{$F$(X90))}&\multicolumn{1}{ M{1.2cm} }{$F$(X180)}  &\multicolumn{1}{ M{1.2cm} }{$F$($\mathcal{I}$)}  &\multicolumn{1}{ M{1.2cm} }{$\epsilon_{\rm oc}$}  &\multicolumn{1}{ M{1.2cm} }{$\epsilon_{\rm oc,in}$}\\\hline
 \multicolumn{1}{  M{1.5cm} }{$T_2^*\sim$ 160 ns} &\multicolumn{1}{ M{1.2cm} }{0.9985} &\multicolumn{1}{ M{1.2cm} }{0.9995} &\multicolumn{1}{ M{1.2cm} }{0.9992}&\multicolumn{1}{ M{1.2cm} }{0.0009}&\multicolumn{1}{ M{1.2cm} }{0.0006}\\ \cline{1-6}
 \multicolumn{1}{  M{1.5cm} }{$T_2^*\sim$ 80 ns} &\multicolumn{1}{ M{1.2cm} }{0.9976} &\multicolumn{1}{ M{1.2cm} }{0.9975} &\multicolumn{1}{ M{1.2cm} }{0.9974} &\multicolumn{1}{ M{1.2cm} }{0.0025} &\multicolumn{1}{ M{1.2cm} }{0.0021}\\ \cline{1-6}
\multicolumn{1}{  M{1.5cm} }{5\% miscal, $T_2^*\sim$ 80 ns} &\multicolumn{1}{ M{1.2cm} }{0.9985}  &\multicolumn{1}{ M{1.2cm} }{0.9974} &\multicolumn{1}{ M{1.2cm} }{0.9972}  &\multicolumn{1}{ M{1.2cm} }{0.0023} &\multicolumn{1}{ M{1.2cm} }{0.0019}\\ \cline{1-6}
\multicolumn{1}{  M{1.5cm} }{-5\% miscal, $T_2^*\sim$ 80 ns} &\multicolumn{1}{ M{1.2cm} }{0.9951} &\multicolumn{1}{ M{1.2cm} }{0.9982}   &\multicolumn{1}{ M{1.2cm} }{0.9935} &\multicolumn{1}{ M{1.2cm} }{0.0044}   &\multicolumn{1}{ M{1.2cm} }{0.0039}\\	\cline{1-6}
\multicolumn{1}{  M{1.5cm} }{Distorted, $T_2^*\sim$ 80 ns} &\multicolumn{1}{ M{1.2cm} }{0.9822} &\multicolumn{1}{ M{1.2cm} }{0.9871}   &\multicolumn{1}{ M{1.2cm} }{0.9615} &\multicolumn{1}{ M{1.2cm} }{0.0230}   &\multicolumn{1}{ M{1.2cm} }{0.0113}\\ \hline\hline
\end{tabular}

\caption[]{Fidelity $F$ (Eq. (1)), $\epsilon_{\rm oc}$ and $\epsilon_{\rm oc,in}$ from numerical simulations ($T_1$ and $T_2$ effects are not considered). $\epsilon_{\rm oc}$ and $\epsilon_{\rm oc,in}$ are the average BEPG and IEPG of the three OC pulses. $F$(Y90) and $F$(Y180) are the same as $F$(X90) and $F$(X180), respectively. `$T_2^*\sim$ 160 ns' and `$T_2^*\sim$ 80 ns' denote the cases when simulating using the Larmor frequency distributions that correspond to the experiments with and without SEL, respectively. `5\% miscal' and `-5\% miscal' denote the cases when simulating OC pulses with powers that are 5\% larger and smaller than the ideal pulse power, respectively. `Distorted' denotes the case where the distorted OC pulses ($\mathcal{T}_{\rm meas}\cdot W$) are simulated. The $T_2^*$ values are estimated from the experimentally measured spectra using a single Lorentzian line shape. In our numerical simulations, we try to more closely match the measured spectra by using Larmor frequency distributions composed of multiple Lorentzian and/or Gaussian line shapes instead of just one.} 
\label{numerical}
\end{table}

%Fitting model for $\langle\sigma_x\rangle^2+\langle\sigma_y\rangle^2+\langle\sigma_z\rangle^2$: $A+Bu^m$, where $u$ is the unitarity, $A$ and $B$ are constants incorporating SPAM %error. The constant offset $A$ is from the identity component. If the measurement is exactly of a traceless observable, $A$ will be identically zero. In our fitting, we constraint $A$ to be zero. One reason is that we don't have enough data points to get a confident fitting of $A+Bu^m$. The other reason is that we believe the noise causing nonzero $A$ is small and won't have significant effect on the fitting of $u$ for the regime we have. We make use of the GST results to simulate very long RB sequences to test the effect of nonzero $A$ in the fitting.

%\begin{figure}[h!]
%%\centering
%\includegraphics[width=0.45\textwidth]{compare_of_Unitarity.eps}
%\caption[]{}
%\label{regular}
%\end{figure}

\begin{figure}[h!]
%%\centering
\includegraphics[width=0.25\textwidth]{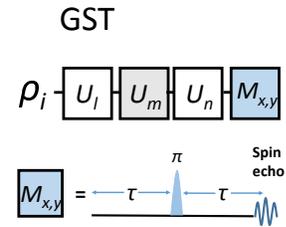}
\caption[]{(Color online)  Quantum circuit for performing GST. In all experiments, the initial state $\rho_i$ is $\sigma_z$ and the 
measurements $M_{x,y}$ are spin echo detection sequences for measuring $\langle\sigma_{x,y}\rangle$. In $M_{x,y}$, the $\pi$ pulse is a 35 ns Gaussian pulse around the \emph{y}-axis, and 
$\tau$=700 ns represents a delay. The $\pi$ pulse acts as the refocusing pulse. The spin echo appears at time $\tau$ after the refocusing pulse and the spin magnetization in the $x$-$y$ plane is detected inductively.}
\label{gst}
\end{figure}

\section{Gate set tomography}

\begin{table}[h!]
	\footnotesize
	\centering
	\begin{tabular}{  M{1.2cm} M{0.6cm}   M{1.6cm} M{1.5cm}  M{1.5cm}  M{1.5cm}  }
		\hhline{~~====}
		& &\multicolumn{1}{ M{1.6cm}  }{Experimental short sequences of RB/PB}&\multicolumn{1}{ M{1.5cm}  }{GST simulation of RB/PB}&\multicolumn{1}{ M{1.5cm}
		}{GST estimates A}&\multicolumn{1}{ M{1.5cm}
		}{GST estimates B}  \\ [2ex]\cline{1-6}
		\multicolumn{1}{   M{1.2cm}  }{\multirow{2}{*}{$\mathcal{T}=\mathcal{T}_{\rm meas}$,}}
		&\multicolumn{1}{ M{0.6cm}  }{$\epsilon$} &\multicolumn{1}{ M{1.5cm}
		}{0.0081(29)}&\multicolumn{1}{ M{1.5cm}  }{$0.0086(1)$}&\multicolumn{1}{ M{1.5cm}  }{$0.0084_{-31}^{+57}$} 
		&\multicolumn{1}{ M{1.5cm}  }{--}\\
		[2ex]
		\multicolumn{1}{   M{1.2cm}  }{SEL} &\multicolumn{1}{ M{0.6cm}
		}{$\epsilon_{{\rm in}}$} &\multicolumn{1}{ M{1.5cm}
		}{0.0078(8)}&\multicolumn{1}{ M{1.5cm}  }{$ 0.0076(0)$}&\multicolumn{1}{ M{1.5cm}  }{$0.0076_{-31}^{+57}$} 
		&\multicolumn{1}{ M{1.5cm}  }{$0.0084^{+57}_{-31}$} 
	\\[2ex]\cline{1-6}
			\multicolumn{1}{   M{1.2cm}  }{\multirow{2}{*}{$\mathcal{T}=\mathcal{T}_{\rm meas}$,}} &\multicolumn{1}{
		M{0.6cm}  }{$\epsilon$} &\multicolumn{1}{ M{1.5cm}
		}{0.0113(33)}&\multicolumn{1}{ M{1.5cm}  
		}{$0.0124(1)$}&\multicolumn{1}{ M{1.5cm}  
		}{$0.0121^{+38}_{-30}$}&\multicolumn{1}{ M{1.5cm}  }{--}\\[2ex]
		\multicolumn{1}{   M{1.2cm}  }{no SEL} &\multicolumn{1}{ M{0.6cm}
		}{$\epsilon_{{\rm in}}$} &\multicolumn{1}{ M{1.5cm}
		}{0.0101(17)}&\multicolumn{1}{ M{1.5cm}  
		}{$0.0111(1)$}&\multicolumn{1}{ M{1.5cm}  
		}{$0.0111^{+38}_{-30}$}&\multicolumn{1}{ M{1.5cm}  
		}{$0.0120^{+38}_{-30}$}
		\\[2ex] \cline{1-6}
		\multicolumn{1}{   M{1.2cm}  }{\multirow{2}{*}{$\mathcal{T}$=1,}}
		&\multicolumn{1}{ M{0.6cm}  }{$\epsilon$} &\multicolumn{1}{ M{1.5cm}
		}{0.0285(105)}&\multicolumn{1}{ M{1.5cm}  }{$0.0331(4)$}&\multicolumn{1}{ M{1.5cm}  }{$0.0331_{-76}^{+151}$} 
		&\multicolumn{1}{ M{1.2cm}  }{--}
		\\[2ex]
		\multicolumn{1}{   M{1.2cm}  }{no SEL} &\multicolumn{1}{
		M{0.6cm}
		}{$\epsilon_{{\rm in}}$} &\multicolumn{1}{ M{1.5cm}
	}{0.0205(78)}&\multicolumn{1}{ M{1.5cm}  }{$0.0247(3)$}&\multicolumn{1}{ M{1.5cm}  }{$0.0250_{-71}^{+150}$}
	&\multicolumn{1}{ M{1.5cm}  }{$0.0331^{+151}_{-76}$}
		\\[2ex]
	\hline\hline
\end{tabular}

\caption[Gate set tomography results]{BEPG $\epsilon$ and IEPG $\epsilon_{\rm in}$ per Clifford gate under different conditions. `$\mathcal{T}$=1, no SEL' denotes the case of unmodified OC pulses; `$\mathcal{T}=\mathcal{T}_{\rm meas}$,  SEL' and `$\mathcal{T}=\mathcal{T}_{\rm meas}$, no SEL' denote the cases of OC pulses  modified by taking into account  $\mathcal{T}_{\rm meas}$ with and without the spin packet selection sequence (SEL), respectively.  The first column  gives the values derived from experimental decays of the RB and PB sequences with 1$\sim$4 Clifford gates. The second column gives the values derived from the simulated decays of the RB and PB sequences with up to 55 Clifford gates, using the process matrices derived in GST experiments. The third and fourth columns give the values calculated from $\mathbf{G}=\mathcal{S}\mathcal{P}\mathcal{Z}$ using the process matrices reconstructed by GST. The error bars in the first and second columns come from the fitting to the models in Eq. (17) of the main text and indicate the 95\% confidence interval. The error bars in the third and fourth columns are calculated using the best and worst process matrices when sampling the parameters of the process matrices within two standard deviations under the CPTP constraints. The $\epsilon_{\rm in}$ in the third and fourth columns are calculated using  $\epsilon_{\rm in}=(1-\sqrt{_{\rm av}u})/2$ and $\epsilon_{\rm in}=(1-\sqrt{u_{\rm av}})/2$, where $_{\rm av}u$ and $u_{\rm av}$ are defined in Eqs. (\ref{eq:avuni}) and (\ref{eq:uniav}), respectively. Both $\epsilon$ and $\epsilon_{\rm in}$ in the third column agree with the values in the second column, indicating the robustness of RB and PB to the realistic gate-dependent noise. The $\epsilon_{\rm in}$ in the fourth column deviate from the values in the second column, indicating PB gives the average of the unitarities ($_{\rm av}u$) instead of the unitarity of the average noise ($u_{\rm av}$).} %correspond to the average of the stochastic errors of the gate-dependent noise in Eq. (\ref{eq:avuni}) and the stochastic error of the noise averaged over the Clifford gates in Eq. (\ref{eq:uniav}), respectively.}
\label{tab:parameters}
\end{table}

The RB and PB protocols assume that the noise has little or no gate dependence, 
which is likely violated in real systems. To study the robustness of RB and PB 
to gate-dependent noise,
we also implement GST \cite{GST1,GST2} to reconstruct the process matrices of the 
OC pulses and use them to  compare with the benchmarking results. 
Compared to standard quantum process tomography (QPT), GST is more robust to state preparation and measurement (SPAM) errors and thus provides 
more accurate information about the target gates. GST is implemented by 
applying three operations
$U_l,U_m,U_n\in$ \{X90, Y90, X180, Y180, $\mathcal{I}$\} to $\rho_i = \sigma_z$ 
and measuring $\langle \sigma_x\rangle$ and $\langle\sigma_y\rangle$ (see Fig. 
\ref{gst}). The Pauli process matrices \cite{Paulitransfer} of the OC 
pulses are then estimated by minimizing the variance between the estimated and 
experimental values of $\langle \sigma_x\rangle$ and $\langle\sigma_y\rangle$ 
\cite{GST1}. Let $\mathcal{Q}^{exp}$ =\{$Q_1$, $Q_2$, $Q_3$, 
$Q_4$, $Q_5$\} denote the Pauli process matrices of the 
experimentally-implemented \{X90,~Y90,~X180,~Y180,~$\mathcal{I}$\} pulses 
and 
$\textbf{m}_{k,lmn}$ denote the expectation value of $\sigma_k$ after 
applying 
$U_l,U_m,U_n$. We estimate the Pauli process matrices by measuring the 
expectation values for each of the 125 combinations of the OC pulses and 
minimizing
\begin{align}
LSQ(\mathcal{Q}) = \sum_{k,l,m,n} |{\bf m}_{k,lmn}-\langle\!\langle
M_k|Q_lQ_mQ_n|\rho_i\rangle\!\rangle|^2,\label{LSQ}\tag{S6}\\
k=x,y;l,m,n=1,\cdots,5\nonumber
\end{align}
under the constraint that the noise is completely-positive and trace-preserving (CPTP) \cite{GST1}. Here, $\langle\!\langle 
M_k|$ and $|\rho_i\rangle\!\rangle$ are the vector forms of the measurement 
operator $M_k$ ($k=x,y$) and initial state $\rho_i$, whose elements are 
$\langle\!\langle M_k|\sigma_j\rangle\!\rangle=\Tr[M_k\sigma_j]$ and 
$\langle\!\langle\sigma_j|\rho_i\rangle\!\rangle=\Tr[\rho_i\sigma_j]/2$ with 
$\sigma_j\in\{\one_2,\sigma_x,\sigma_y,\sigma_z\}$ ($j=1,2,3,4$).

\begin{figure}

\includegraphics[width=0.9\linewidth]{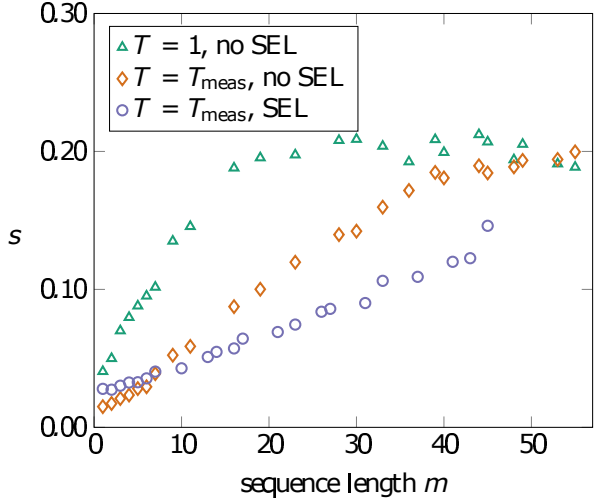}
	\caption[]{(Color online) Standard deviation $s$ between experimental RB results and GST simulations of the expectation values  over 150 random sequences. The results of the three conditons, using the unmodified OC pulses, the transfer function modified OC pulses  and the transfer function modified  OC pulses with SEL, are presented by green triangles, orange  diamonds, and purple circles, respectively. The increase of $s$ with the sequence length can be attributed to the fact that GST was performed using sequences of three 
pulses, which limits the accuracy. Up to the first few gates, the GST simulations agree closely with each individual sequence of the RB experiments for the modified OC pulses, but demonstrate substantial disagreement for the unmodified pulses. We attribute this disagreement to inaccuracies of the GST results in characterizing the large coherent errors in the individual gates resulting from the uncorrected distortion. For the OC pulses modified by the transfer function with SEL, the $s$ values are unexpectedly higher (compared to no SEL) for the first few gates. We attribute this mainly to a smaller signal-to-noise ratio in the case of SEL. Note that the roughly linear increases in $s$ versus sequence length suggest that discrepancies are accumulating stochastically rather than coherently.}%stochastic rather than coherent errors, as mentioned in the main text.} %Up to the first few gates, the case of the OC pulses modified by the transfer function with SEL has larger $s$ than the the case without SEL due to the smaller signal-to-noise ratio in the former case.  due to inaccuracies in characterizing the large coherent errors  in the individual gates.}
	\label{GSTRBdiff}
\end{figure}

The estimated process matrices for the OC pulses are given in the 
next section and their fidelities are given in Table \ref{GST}. We then use these measured process matrices to construct the process matrices of Clifford gates from $\mathbf{G}=\mathcal{S}\mathcal{P}\mathcal{Z}$ ($\mathcal{S}\in 
\{\rm \mathcal{I},X90,Y90$\}, $\mathcal{P}\in \{\rm \mathcal{I},Y180\}$,
and $\mathcal{Z}\in \{\rm \mathcal{I},Z90,Z180,Z270\}$) and calculate the benchmarking average error per gate (BEPG) $\epsilon$, which are listed in the `GST estimates A' column of Table \ref{tab:parameters}. We found that the calculated $\epsilon$ values are closer to the fitting results of 
the first few gates of the experimental
data within error bars than the fitting results obtained by including all gates in the RB and PB sequences ({\it i.e.} values listed in Table I of the main text). We believe this is due to the GST sequences containing only three gates, which are not appropriate to predict the asymptotic behaviour of the RB and PB sequences (also see Fig. \ref{GSTRBdiff}). 

%We also 
%plot the difference between the GST and experimental estimates per sequence as 
%a function of the sequence length. The linear increase with the sequence 
%length 
%can be attributed to the fact that GST was performed using sequences of three 
%pulses, which limits the accuracy. The substantial disagreement between the 
%GST 
%simulations and the experimental data per sequence for the undistorted OC 
%pulses is primarily due to inaccuracies in characterizing the rotation axis of 
%the substantial coherent errors. However, the GST simulations of the 
%undistorted OC pulses still agree relatively well on average, as RB and PB are 
%both insensitive to the rotation axis of a coherent error (RB returns the 
%average gate error, which only depends upon the magnitude of a coherent error, 
%and PB returns the unitarity, which is designed to be invariant under unitary 
%rotations).

We also use two different methods to calculate the incoherent error per gate (IEPG) $\epsilon_{in}$ using the process matrices derived from GST. The PB protocol has only been analyzed under the assumption that the noise is gate-independent. One immediate question is whether the decay parameter of PB should be the
average of the unitarities of the gate-dependent noise,
\begin{align}\label{eq:avuni}\tag{S7}
_{\rm av}u = \lvert \mathbf{G}\rvert^{-1}\sum_{G\in\mathbf{G}}u[\mathcal{E}(G)],
\end{align}
or the unitarity of the average noise,
\begin{align}\label{eq:uniav}\tag{S8}
u_{\rm av} = u\left[\lvert
\mathbf{G}\rvert^{-1}\sum_{G\in\mathbf{G}}\mathcal{E}(G)\right].
\end{align}
These two quantities are not equal in general, as the unitarity is not a linear
function of quantum channels. The IEPG for both quantities are listed in columns `GST estimates A' and `GST estimates B' of Table~\ref{tab:parameters}. Although the table shows that $_{\rm av}u$ is closer to the experimentally observed decay than $u_{\rm av}$, the error bars are large and we cannot get statistically meaningful conclusion whether the PB protocol measures the average of the incoherent errors of the gate-dependent noise or the incoherent error of the noise averaged over the gates being benchmarked.
%\begin{table}[h!]
%\scriptsize
%\centering
%\begin{tabular}{l C{24mm} C{24mm} C{24mm}}
%	\hline\hline
%	& $\mathcal{T}=1$ & $\mathcal{T}=\mathcal{T}_{\rm meas}$ & 
%	$\mathcal{T}=\mathcal{T}_{\rm meas}$ \\
%	& without SEL & without SEL & with SEL \\\hline
%$\epsilon$ & 0.0234$\pm$0.0011 & 0.0073$\pm$0.0002 &0.0063$\pm$0.0001  \\
%$\epsilon_{\rm in}$ & 0.0105$\pm$0.0010 & 0.0066$\pm$0.0002 &0.0054$\pm$0.0002  
%\\
%$\epsilon_\diamond$ & \\
%	\hline\hline
%\end{tabular}
%\caption[Results]{Estimates of the A-EPG $\epsilon$, I-EPG $\epsilon_{\rm in}$ 
%and W-EPG $\epsilon_\diamond$ per Clifford gate with OC pulses that assume a 
%flat resonator transfer function ($\mathcal{T}=1$) or are distorted based on 
%the measured transfer function ($\mathcal{T} = \mathcal{T}_{\rm meas}$), and 
%with or without spin packet selection sequences (SEL) respectively.}  
%\label{tab:parameters}
%\end{table}

Therefore, we use the help of simulation to further study the performance of PB and RB. We use the experimental GST process matrices as realistic noise models to simulate RB and PB with the same sequences used in the experiment (up to 55 Clifford gates). The derived decay rates of the simulations are listed in the column `GST simulation of RB/PB' of Table \ref{tab:parameters}. The values in the `GST simulation of RB/PB' and in the `GST es-
timates A' agree to within the fitting uncertainty (note that the uncertainties
in the `GST estimates A' and `GST estimates B' are not relevant as our
simulations are of the average GST reconstructions), indicating that under realistic
noise models consistent with our system, both the RB and PB work very
well in estimating the BEPG and IEPG. Furthermore, our simulations suggest
the PB protocol measures the average of the incoherent errors of the gate-dependent noise, which
is sensitive to individual calibration errors (even if these gate-dependent
calibration errors average into an incoherent process).
\section{Gate set tomography results}
Here we list the reconstructed Pauli transfer matrices obtained from the GST experiments. There are three cases: (i) OC pulses are modified by $\mathcal{T}_{\rm meas}$ and SEL is used; (ii) OC pulses are modified by $\mathcal{T}_{\rm meas}$ and no SEL; (iii) unmodified pulses and no SEL. 

In case (i): 
\begin{align}
\rm{X90}=\begin{bmatrix}
    1  & 0 & 0 & 0\\
   -0.0037  &  0.9886  & -0.0249  &  0.0112\\
   -0.0051  & 0.0102   & 0.0451   &-0.9899\\
   0.0021  & 0.0204   & 0.9857  & 0.0495
\end{bmatrix},\nonumber
\end{align}
\begin{align}
\rm{Y90}=\begin{bmatrix}
    1  & 0 & 0 & 0\\
   0.0053  &  0.0511  & -0.0152 &   0.9925 \\
   0.0035  &  0.0163  &  0.9947  &  0.0121\\
    0.0011  & -0.9939  &  0.0167  &  0.0510
\end{bmatrix},\nonumber
\end{align}
\begin{align}
\rm{X180}=\begin{bmatrix}
    1  & 0 & 0 & 0\\
   -0.0054 &  0.9807  & 0.0052 & -0.0137 \\
   -0.0060 & 0.0023  & -0.9886   & 0.0253\\
   -0.0070 & -0.0194 &  -0.0200  &-0.9862
\end{bmatrix},\nonumber
\end{align}
\begin{align}
\rm{Y180}=\begin{bmatrix}
    1  & 0 & 0 & 0\\
   0.0078 & -0.9890 &  -0.0039   &-0.0300 \\
   0.0023 & -0.0034 &  0.9873  & -0.0107\\
   -0.0023  & 0.0230  & -0.0121 & -0.9830
\end{bmatrix},\nonumber
\end{align}
\begin{align}
\rm{\mathcal{I}}=\begin{bmatrix}
    1  & 0 & 0 & 0\\
   -0.0014  & 0.9784  & 0.0774  & 0.0490 \\
    0.0018  & -0.0663 &  0.9746 & -0.0709\\
   0.0029 & -0.0587  & 0.0711  & 0.9809
\end{bmatrix}.\nonumber
\end{align}
In case (ii): 
\begin{align}
\rm{X90}=\begin{bmatrix}
    1  & 0 & 0 & 0\\
   -0.0001 &   0.9831  & -0.0316 &  -0.0112 \\
    -0.0011 & -0.0115  & 0.0323  & -0.9752\\
    -0.0025 &   0.0375 &  0.9904  & 0.0326
\end{bmatrix},\nonumber
\end{align}
\begin{align}
\rm{Y90}=\begin{bmatrix}
    1  & 0 & 0 & 0\\
   -0.0001  &  0.0324  &  0.0130  &  0.9793 \\
   0.0008  &  0.0359   & 0.9875  & -0.0112\\
   0.0009  & -0.9887   & 0.0454  &  0.0334
\end{bmatrix},\nonumber
\end{align}
\begin{align}
\rm{X180}=\begin{bmatrix}
    1  & 0 & 0 & 0\\
   -0.0006  &  0.9760  &  0.0284  & -0.0132 \\
   0.0027  &  0.0242  & -0.9833   & 0.0115\\
   -0.0015  & -0.0096 &   0.0011  & -0.9907
\end{bmatrix},\nonumber
\end{align}
\begin{align}
\rm{Y180}=\begin{bmatrix}
    1  & 0 & 0 & 0\\
   0.0025 &  -0.9858  & -0.0211 &  -0.0139\\
   -0.0017 &  -0.0248 &   0.9835  & -0.0045\\
   0.0001  &  0.0052  & -0.0117  & -0.9853
\end{bmatrix},\nonumber
\end{align}
\begin{align}
\mathcal{I}=\begin{bmatrix}
    1  & 0 & 0 & 0\\
   -0.0013 &  0.9703 &  0.0627  & 0.0553\\
    0.0004 & -0.0489 &  0.9639  & -0.1045\\
   -0.0014 &  -0.0662  &  0.1064  & 0.9687
\end{bmatrix}.\nonumber
\end{align}
In case (iii): 
\begin{align}
\rm{X90}=\begin{bmatrix}
    1  & 0 & 0 & 0\\
   -0.0022 &  0.9557  &  0.1418 &  -0.1359 \\
   0.0002 & -0.1352  & -0.0813  & -0.9685\\
   0.0061  & -0.1594  &  0.9464 &  -0.0679
\end{bmatrix},\nonumber
\end{align}
\begin{align}
\rm{Y90}=\begin{bmatrix}
    1  & 0 & 0 & 0\\
   0.0029 & -0.0743  &  0.1025  &  0.9628\\
   -0.0059  & -0.1145 &  0.9502  & -0.0976\\
    0.0024  & -0.9610  & -0.1246  & -0.0932
\end{bmatrix},\nonumber
\end{align}
\begin{align}
\rm{X180}=\begin{bmatrix}
    1  & 0 & 0 & 0\\
   0.0044 & 0.9531  & -0.0016  &  0.1211\\
   0.0036  & -0.0126 &  -0.9621  &  0.0081\\
   -0.0208  &  0.1438 &  0.0013 & -0.9624
\end{bmatrix},\nonumber
\end{align}
\begin{align}
\rm{Y180}=\begin{bmatrix}
    1  & 0 & 0 & 0\\
   0.0016 &  -0.9598   & 0.0121   &-0.0085\\
  -0.0001  &  0.0083   & 0.9462  &  0.1687\\
  -0.0027  & -0.0234   & 0.1564  & -0.9575
\end{bmatrix},\nonumber
\end{align}
\begin{align}
\rm{\mathcal{I}}=\begin{bmatrix}
    1  & 0 & 0 & 0\\
   -0.0044  &  0.9029  & -0.0594 &  -0.0698\\
   -0.0002  & -0.0180  &  0.9424 &   0.0190\\
   -0.0183  &  0.0563  & -0.0120 &   0.9083
\end{bmatrix}.\nonumber
\end{align}

\section{Design of Benchmarking sequences and Simulation Using Lindblad Equation}

We generated 150 random sequences for each length of Clifford gates  for the RB and PB experiments, up to 55 Clifford gates. Each sequence had randomly selected Clifford gates such that the Clifford group was sampled uniformly.  For the experiment, we only used Clifford gate sequences of length $m$=1,$\ldots$, 7, 9, 11, 16, 19, 23, 28, 30, 33, 36, 39, 40, 44, 45, 48, 49, 53, 55 for the RB without SEL; length $m$=1,$\ldots$, 7, 9, 14, 17, 20, 22, 25, 27, 31, 35, 37, 40, 41, 42, 45, 49, 54, 55 for the PB without SEL; length $m$=1,$\ldots$, 7, 10, 13, 14, 16, 17, 21, 23, 26, 27, 31, 33, 37, 41, 43, 45 for the RB with SEL; length $m$=1,$\ldots$, 8,  9, 13, 17, 18, 19, 22, 27, 29, 32, 34, 38, 40, 43, 47 for the PB with SEL. Longer sequences could not be sampled mainly due to the limitation of the TWT amplifier.

We also numerically simulate our benchmarking experiments.  The simulation evolves an initial density matrix according to the Linblad model using the same pulse sequences as in the experiments. Experimentally measured $T_1$=160 $\operatorname{\mu s}$ and $T_2$=30  $\operatorname{\mu s}$ are used in the simulation. We incorporate the effects of $T_2^*$ and local $B_1$ field inhomogeneities by averaging the simulation over these distributions (see Fig.~\ref{spectrum} for the Larmor frequency distribution, which corresponds to the $T_2^*$ effect, and see Fig. \ref{B1} for the local $B_1$ distribution). The SEL sequence is used in the experiments to effectively extend $T_{2, {\rm no \; SEL}}^*\sim$80 ns.  This sequence is comprised of eight 2$\pi$ rotation pulses, each separated by an order of $T_2$ to allow the transverse components to dephase \cite{ESRRB}.  This SEL sequence selects a subset of the spin ensemble within narrower Larmor frequency distribution while dephasing most of the off-resonance spin packets as shown in Fig.~\ref{spectrum}, but we found that it does not have much effect on the $B_1$ distribution. The SEL step is incorporated in the simulation by using a distribution that matches the Larmor frequency distribution measured in the experiments with SEL sequences ($T_{2, {\rm SEL}}^*\sim$160 ns).% an extended value of $T_{2, {\rm SEL}}^*\sim$160 ns \cite{T2star} to match the Larmor frequency distribution measured in the experiments with SEL sequences. 

\begin{figure}[h!]
\includegraphics[width=0.45\textwidth]{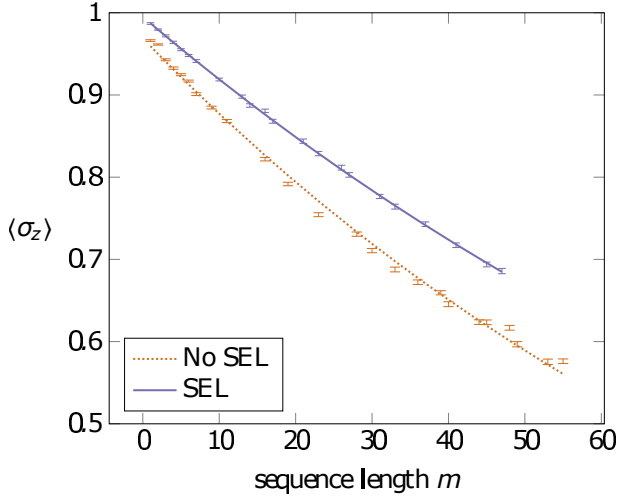}
\caption[]{(Color online) Simulation results of RB using the same pulse sequences as in the experiments. `SEL' and `No SEL' denote the cases of $T_2^*$ $\sim$ 160 ns and $T_2^*$ $\sim$ 80 ns, respectively. The solid and dotted lines are least squares fits to $B(1-2\epsilon)^m$ where $\epsilon$ is the BEPG for the Clifford group.  The error bars indicate the standard 
	error of the mean.}
\label{fig_rb_sim}
\end{figure}

\begin{figure}[h!]
\includegraphics[width=0.45\textwidth]{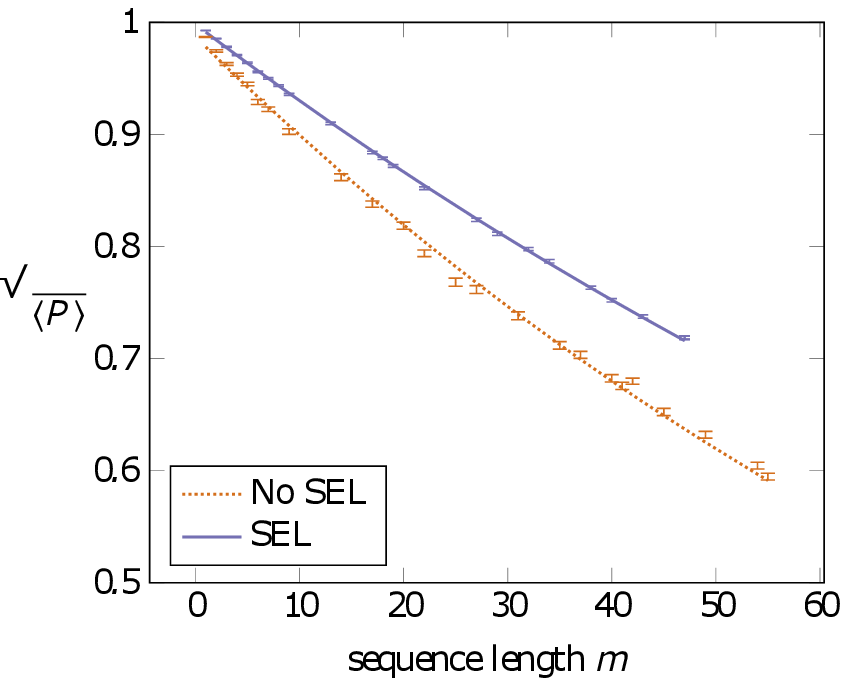}
\caption[]{(Color online) Simulation results of PB using the same pulse sequences as in the experiments. `SEL' and `No SEL' denote the cases of $T_2^*$ $\sim$ 160 ns and $T_2^*$ $\sim$ 80 ns, respectively. The solid and dotted lines are least squares fits to $\sqrt{B'}(1-2\epsilon_{\rm in})^{(m-1)}$ where $\epsilon_{\rm in}$ is the IEPG for the Clifford group. The error bars indicate the standard 
	error of the mean.}
\label{fig_unit_sim}
\end{figure}

Figures \ref{fig_rb_sim} and \ref{fig_unit_sim} show the simulation results of RB and PB protocols when taking and not taking SEL sequences into account. These simulations only take into account $T_2^*$ distribution since we simulated that our measured $B_1$ field inhomogeneity has little effect on our BEPG ($< 10^{-4}$). The values from the simulation with no SEL sequences are $\epsilon = 0.0049(1)$ and $\epsilon_{\rm in} = 0.0046(1)$.  The values from the simulation using the  Larmor frequency distribution with SEL sequences are $\epsilon =0.0040(0)$ and $\epsilon_{\rm in} = 0.0035(0)$. As simulations assume no pulse distortions (\textit{i.e}., perfect knowledge of $\mathcal{T}$), the only coherent error source in the simulations is the imperfection in the OC pulse design which is very small (see Table \ref{numerical}).% (GRAPE pulses are calculated to have a fidelity $\geq 0.995$).

In the case with short $T_2^*$, the simulated decays slightly deviate from a single exponential decay, while the simulated decays in the case with long $T_2^*$ can be fitted very well to a single exponential decay up to $\sim$ 50 gates, which agrees with the experimental results. This indicates the $T_2^*$ effect contributes to the non-Markovian noise.

%\vspace{12pt}

%\begin{thebibliography}{10}

%\bibitem{ESRRBs} D. K. Park, G. Feng, R. Rahimi, J. Baugh, and R. Laflamme, Journal of Magnetic Resonance 267, 68–78 (2016).

%\bibitem{khaneja2005optimals}
%N.~Khaneja, T.~Reiss, C.~Kehlet, T.~Schulte-Herbr{\"u}ggen, and S.~J. Glaser, Journal of Magnetic Resonance 172~(2),
%  296--305 (2005).
  
%\bibitem{T2star} The $T_2^*$ values are estimated from the measured spectra using a single lorentzian line shape. In our simulations,  we try to more closely match the measured spectra by using %Larmor frequency distributions composed of multiple Lorentzian and/or Gaussian line shapes instead of just one.%we tried to mimic the measured distributions as closely as possible instead of using single lorentzian line shape.  

%\bibitem{GST1s} S. T. Merkel, J. M. Gambetta, J. A. Smolin, S.
%Poletto, A. D. C\'{o}rcoles, B. R. Johnson, C. A. Ryan, and M.
%Steffen,  Phys. Rev. A 87, 062119 (2013).

%\bibitem{GST2s} R. Blume-Kohout, J. K. Gamble, E. Nielsen, J.
%Mizrahi, J. D. Sterk, and P. Maunz, arXiv:1310.4492 (2013).

%\bibitem{Paulitransfer} J. M. Chow \textit{et al.}, Phys. Rev. Lett. 109, 060501 (2012).

%\expandafter\ifx\csname url\endcsname\relax
%  \def\url#1{\texttt{#1}}\fi
  %
%\expandafter\ifx\csname urlprefix\endcsname\relax\def\urlprefix{URL }\fi
%\expandafter\ifx\csname href\endcsname\relax
 % \def\href#1#2{#2} \def\path#1{#1}\fi

%\end{thebibliography}

\end{document}